%% file: hubbard1.tex
\overfullrule=0pt

\newcount\mgnf  
\mgnf=0

\ifnum\mgnf=0
\def\openone{\leavevmode\hbox{\ninerm 1\kern-3.3pt\tenrm1}}%
\def\*{\vglue0.3truecm}\fi
\ifnum\mgnf=1
\def\openone{\leavevmode\hbox{\ninerm 1\kern-3.63pt\tenrm1}}%
\def\*{\vglue0.5truecm}\fi

\ifnum\mgnf=0
   \magnification=\magstep0
   \hsize=15truecm\vsize=23.truecm
   \parindent=4.pt\baselineskip=0.45cm
\font\titolo=cmbx12 \font\titolone=cmbx10 scaled\magstep 2
\font\cs=cmcsc10  \font\ottorm=cmr8
 \font\euftw=eufm10 \font\msytw=msbm10
\font\msytww=msbm8  \font\indbf=cmbx10
scaled\magstep1  \fi \ifnum\mgnf=1
   \magnification=\magstep1\hoffset=0.truecm
   \hsize=14truecm\vsize=24.truecm
   \baselineskip=18truept plus0.1pt minus0.1pt \parindent=0.9truecm
   \lineskip=0.5truecm\lineskiplimit=0.1pt      \parskip=0.1pt plus1pt
\font\titolo=cmbx12 scaled\magstep 1 \font\titolone=cmbx10
scaled\magstep 3 \font\cs=cmcsc10 scaled\magstep 1
\font\ottorm=cmr8 scaled\magstep 1 
scaled\magstep1 \font\euftw=eufm10 scaled\magstep1
\font\msytw=msbm10 scaled\magstep1 \font\msytww=msbm8
scaled\magstep1 
 \font\indbf=cmbx10
scaled\magstep2 \fi

\global\newcount\numsec\global\newcount\numapp
\global\newcount\numfor\global\newcount\numfig\global\newcount\numsub
\global\newcount\numlemma\global\newcount\numtheorem\global\newcount\numdef
\global\newcount\appflag \numsec=0\numapp=0\numfig=1
\def\veroparagrafo{\number\numsec}\def\veraformula{\number\numfor}
\def\veraappendice{\number\numapp}\def\verasub{\number\numsub}
\def\verafigura{\number\numfig}
\def\verolemma{\number\numlemma}
\def\verotheorem{\number\numtheorem}
\def\veradef{\number\numdef}

\def\section(#1,#2){\advance\numsec by 1\numfor=1\numsub=1%
\numlemma=1\numtheorem=1\numdef=1\appflag=0%
\SIA p,#1,{\veroparagrafo} %
\write15{\string\Fp (#1){\secc(#1)}}%
\write16{ sec. #1 ==> \secc(#1)  }%
\hbox to \hsize{\titolo\hfill \number\numsec. #2\hfill%
\expandafter{\alato(sec. #1)}}\*}

\def\appendix(#1,#2){\advance\numapp by 1\numfor=1\numsub=1%
\numlemma=1\numtheorem=1\numdef=1\appflag=1%
\SIA p,#1,{A\veraappendice} %
\write15{\string\Fp (#1){\secc(#1)}}%
\write16{ app. #1 ==> \secc(#1)  }%
\hbox to \hsize{\titolo\hfill Appendix A\number\numapp. #2\hfill%
\expandafter{\alato(app. #1)}}\*}

\def\senondefinito#1{\expandafter\ifx\csname#1\endcsname\relax}

\def\SIA #1,#2,#3 {\senondefinito{#1#2}%
\expandafter\xdef\csname #1#2\endcsname{#3}\else \write16{???? ma
#1#2 e' gia' stato definito !!!!} \fi}

\def \Fe(#1)#2{\SIA fe,#1,#2 }
\def \Fp(#1)#2{\SIA fp,#1,#2 }
\def \Fg(#1)#2{\SIA fg,#1,#2 }
\def \Fl(#1)#2{\SIA fl,#1,#2 }
\def \Ft(#1)#2{\SIA ft,#1,#2 }
\def \Fd(#1)#2{\SIA fd,#1,#2 }

\def\etichetta(#1){%
\ifnum\appflag=0(\veroparagrafo.\veraformula)%
\SIA e,#1,(\veroparagrafo.\veraformula) \fi%
\ifnum\appflag=1(A\veraappendice.\veraformula)%
\SIA e,#1,(A\veraappendice.\veraformula) \fi%
\global\advance\numfor by 1%
\write15{\string\Fe (#1){\equ(#1)}}%
\write16{ EQ #1 ==> \equ(#1)  }}

\def\getichetta(#1){Fig. \verafigura%
\SIA g,#1,{\verafigura} %
\global\advance\numfig by 1%
\write15{\string\Fg (#1){\graf(#1)}}%
\write16{ Fig. #1 ==> \graf(#1) }}

\def\etichettap(#1){%
\ifnum\appflag=0{\veroparagrafo.\verasub}%
\SIA p,#1,{\veroparagrafo.\verasub} \fi%
\ifnum\appflag=1{A\veraappendice.\verasub}%
\SIA p,#1,{A\veraappendice.\verasub} \fi%
\global\advance\numsub by 1%
\write15{\string\Fp (#1){\secc(#1)}}%
\write16{ par #1 ==> \secc(#1)  }}

\def\etichettal(#1){%
\ifnum\appflag=0{\veroparagrafo.\verolemma}%
\SIA l,#1,{\veroparagrafo.\verolemma} \fi%
\ifnum\appflag=1{A\veraappendice.\verolemma}%
\SIA l,#1,{A\veraappendice.\verolemma} \fi%
\global\advance\numlemma by 1%
\write15{\string\Fl (#1){\lm(#1)}}%
\write16{ lemma #1 ==> \lm(#1)  }}

\def\etichettat(#1){%
\ifnum\appflag=0{\veroparagrafo.\verotheorem}%
\SIA t,#1,{\veroparagrafo.\verotheorem} \fi%
\ifnum\appflag=1{A\veraappendice.\verotheorem}%
\SIA t,#1,{A\veraappendice.\verotheorem} \fi%
\global\advance\numtheorem by 1%
\write15{\string\Ft (#1){\thm(#1)}}%
\write16{ th. #1 ==> \thm(#1)  }}

\def\etichettad(#1){%
\inum\appflag=0{\veroparagrafo.\veradef}%
\SIA d,#1,{\veroparagrafo.\veradef} \fi%
\inum\appflag=1{A\veraappendice.\veradef}%
\SIA d,#1,{A\veraappendice.\veradef} \fi%
\global\advance\numdef by 1%
\write15{\string\Fd (#1){\defz(#1)}}%
\write16{ def. #1 ==> \defz(#1)  }}

\def\Eq(#1){\eqno{\etichetta(#1)\alato(#1)}}
\def\eq(#1){\etichetta(#1)\alato(#1)}
\def\eqg(#1){\getichetta(#1)\alato(fig #1)}
\def\sub(#1){\0\palato(p. #1){\bf \etichettap(#1)\hskip.3truecm}}
\def\lemma(#1){\0\palato(lm #1){\cs Lemma \etichettal(#1)\hskip.3truecm}}
\def\theorem(#1){\0\palato(th #1){\cs Theorem \etichettat(#1)%
\hskip.3truecm}}
\def\definition(#1){\0\palato(df #1){\cs Definition \etichettad(#1)%
\hskip.3truecm}}

\def\equv(#1){\senondefinito{fe#1}$\clubsuit$#1%
\write16{eq. #1 non e' (ancora) definita}%
\else\csname fe#1\endcsname\fi}
\def\grafv(#1){\senondefinito{fg#1}$\clubsuit$#1%
\write16{fig. #1 non e' (ancora) definito}%
\else\csname fg#1\endcsname\fi}

\def\secv(#1){\senondefinito{fp#1}$\clubsuit$#1%
\write16{par. #1 non e' (ancora) definito}%
\else\csname fp#1\endcsname\fi}

\def\lmv(#1){\senondefinito{fl#1}$\clubsuit$#1%
\write16{lemma #1 non e' (ancora) definito}%
\else\csname fl#1\endcsname\fi}

\def\thmv(#1){\senondefinito{ft#1}$\clubsuit$#1%
\write16{th. #1 non e' (ancora) definito}%
\else\csname ft#1\endcsname\fi}

\def\defzv(#1){\senondefinito{fd#1}$\clubsuit$#1%
\write16{def. #1 non e' (ancora) definito}%
\else\csname fd#1\endcsname\fi}

\def\equ(#1){\senondefinito{e#1}\equv(#1)\else\csname e#1\endcsname\fi}
\def\graf(#1){\senondefinito{g#1}\grafv(#1)\else\csname g#1\endcsname\fi}
\def\secc(#1){\senondefinito{p#1}\secv(#1)\else\csname p#1\endcsname\fi}
\def\lm(#1){\senondefinito{l#1}\lmv(#1)\else\csname l#1\endcsname\fi}
\def\thm(#1){\senondefinito{t#1}\thmv(#1)\else\csname t#1\endcsname\fi}
\def\defz(#1){\senondefinito{d#1}\defzv(#1)\else\csname d#1\endcsname\fi}
\def\sec(#1){{\S\secc(#1)}}

\def\BOZZA{
\def\alato(##1){\rlap{\kern-\hsize\kern-1.2truecm{$\scriptstyle##1$}}}
\def\palato(##1){\rlap{\kern-1.2truecm{$\scriptstyle##1$}}}
}

\def\alato(#1){}
\def\galato(#1){}
\def\palato(#1){}

{\count255=\time\divide\count255 by 60
\xdef\hourmin{\number\count255}
        \multiply\count255 by-60\advance\count255 by\time
   \xdef\hourmin{\hourmin:\ifnum\count255<10 0\fi\the\count255}}

\def\oramin{\hourmin }

\def\data{\number\day/\ifcase\month\or gennaio \or febbraio \or marzo \or
aprile \or maggio \or giugno \or luglio \or agosto \or settembre
\or ottobre \or novembre \or dicembre \fi/\number\year;\ \oramin}
\setbox200\hbox{$\scriptscriptstyle \data $}
\footline={\rlap{\hbox{\copy200}}\tenrm\hss \number\pageno\hss}

\let\a=\alpha \let\b=\beta  \let\g=\gamma     \let\d=\delta  \let\e=\varepsilon
  \let\h=\eta    \def\th{\theta}
   
\let\m=\mu    \let\n=\nu            \let\p=\pi      \let\r=\rho
\let\s=\sigma \let\t=\tau        \let\c=\chi
   \let\o=\omega 
 \let\D=\Delta     \let\L=\Lambda  
           
\let\O=\Omega 

\def\\{\hfill\break} \let\==\equiv

\let\io=\infty 

\let\0=\noindent

\let\dpr=\partial 
\let\bs=\backslash

\def\defin{{\buildrel def\over=}}
\def\tende#1{\,\vtop{\ialign{##\crcr\rightarrowfill\crcr
 \noalign{\kern-1pt\nointerlineskip}
 \hskip3.pt${\scriptstyle #1}$\hskip3.pt\crcr}}\,}
\def\otto{\,{\kern-1.truept\leftarrow\kern-5.truept\to\kern-1.truept}\,}
\def\fra#1#2{{#1\over#2}}

\def\PP{{\cal P}}\def\EE{{\cal E}}\def\VV{{\cal V}}
\def\WW{{\cal W}}
\def\TT{{\cal T}}\def\NN{{\cal N}}\def\BB{{\cal B}}
\def\RR{{\cal R}}\def\LL{{\cal L}}
\def\DD{{\cal D}}\def\SS{{\cal S}}

\def\T#1{{#1_{\kern-3pt\lower7pt\hbox{$\widetilde{}$}}\kern3pt}}
\def\VVV#1{{\underline #1}_{\kern-3pt
\lower7pt\hbox{$\widetilde{}$}}\kern3pt\,}
\def\W#1{#1_{\kern-3pt\lower7.5pt\hbox{$\widetilde{}$}}\kern2pt\,}

  \def\sign{{\rm sign}\,}
\def\indica{\leaders \hbox to 0.5cm{\hss.\hss}\hfill}
\def\guida{\leaders\hbox to 1em{\hss.\hss}\hfill}
\mathchardef\oo= "0521

\def\pp{{\bf p}}\def\qq{{\bf q}}\def\xx{{\bf x}}
\def\yy{{\bf y}}\def\kk{{\bf k}}
\def\zz{{\bf z}}
 \def\bP{{\bf P}}\def\rr{{\bf r}}

\def\bT{{\bf T}}

\def\Halmos{\hfill\vrule height6pt width4pt depth2pt \par\hbox to \hsize{}}
\def\virg{\quad,\quad}

\def\ss{{\underline \sigma}}\def\oo{{\underline \omega}}

\def\un{{\underline \nu}}

\def\qed{\raise1pt\hbox{\vrule height5pt width5pt depth0pt}}

 \def\bh{{\bar h}}  

\def\indic{\hbox{\raise-2pt \hbox{\indbf 1}}}
\def\bk#1#2{\bar\kk_{#1#2}}

\def\MMM{\hbox{\euftw M}} 
\def\RRR{\hbox{\msytw R}}

\def\zzzz{\hbox{\msytww Z}}

%
%
%
\def\ins#1#2#3{\vbox to0pt{\kern-#2 \hbox{\kern#1 #3}\vss}\nointerlineskip}
%
%
%
\newdimen\xshift \newdimen\xwidth \newdimen\yshift

\def\insertplot#1#2#3#4#5{\par%
\xwidth=#1 \xshift=\hsize \advance\xshift by-\xwidth \divide\xshift by 2%
\yshift=#2 \divide\yshift by 2%
\line{\hskip\xshift \vbox to #2{\vfil%
#3 \includegraphics{#4.pst}}\hfill \raise\yshift\hbox{#5} }}

\openin14=\jobname.aux \ifeof14 \relax \else
\input \jobname.aux \closein14 \fi
\openout15=\jobname.aux

{\baselineskip=12pt
\centerline{\titolone Rigorous proof of Luttinger liquid}
\centerline{\titolone behavior in the 
1d Hubbard model
}
\vskip1.truecm
\centerline{\titolo{Vieri Mastropietro}}
\vskip.5cm
\centerline{Dipartimento di Matematica, Universit\`a di Roma ``Tor
Vergata''}
\centerline{Via della Ricerca Scientifica, I-00133, Roma}
\*
\*
\vskip1cm \0{\cs Abstract.} {\it 
We give the first rigorous (non perturbative) proof
of Luttinger liquid behavior in the one dimensional Hubbard
model, for small repulsive interaction and values
of the density different from half filling. 
The analysis is based on the combination
of multiscale analysis with Ward  
identities bases on a hidden and approximate local chiral gauge invariance. 
No use is done of exact solutions or
special integrability properties of the Hubbard model, 
and the results can be in fact easily generalized to include 
non local interactions, magnetic 
fields or interaction with external potentials.}
\*
{\cs Key words} {\rm Interacting fermions, spin, Ward identities, Renormalization Group}
\*
\section(1,Introduction)
\* 
\sub(1.1) {\it Historical remarks.}

The Hubbard model describes electrons in a crystalline lattice,
hopping from one site of a lattice to another and interacting
by a repulsive (Coulomb) force with coupling $U>0$. As pointed out by Lieb [L],
the Hubbard model in the theory of interacting electrons has the same role of the Ising
model in the problem of spin-spin correlations, that is it is the simplest model 
displaying many real word features: it is however much more difficult to analyze. 
It is believed that the Hubbard model 
gives a correct description of the properties of many metals
due to the interactions between conduction electrons:
for instance the phenomenon of Mott transition, the anomalous properties of 
high $T_c$ superconductors or the singular properties of quantum wires. 
However the mathematical complexity
of the computations is such that 
this belief is still
far from be substantiated by solid arguments.
While our understanding of the Hubbard model in higher dimensions
is really poor, the situation is of course better in $d=1$;
the interest in such a case is not purely academic as 
in this case the model
is believed to furnish an accurate description of real systems like quantum wires.

In $d=1$ the Hubbard model
can be solved exactly by an extension of the Bethe ansatz,
as it was shown by Lieb and Wu [LW] in the sixties;  it was found that 
in the half filled band case a gap opens in the spectrum, so that 
the system is an insulating, while in the other cases there is no gap
and the system is a metal.
Subsequently many thermodinamical quantities were obtained, see 
for instance [O], [T]; for a review of exact results
see [L].

The exact solution is however 
essentially of no utility for computing 
the correlations, which are the quantities
more directly related to physical observables; 
even if one has the full form of
the wave functions (what is actually {\it not} the case, as the Bethe ansatz
gives them as the solutions of complicate integral equations), 
computing the correlations from them   
is essentially impossible. In particular
an important question which
cannot be answered by the exact solution is if the Hubbard model
is a {\it Fermi liquid} or a {\it Luttinger liquid}. 
The notion of Luttinger liquid was introduced by 
Haldane [H] in the early eighties. While a Fermi liquid is an interacting fermionic
system whose low energy behavior is close to the one of the free
Fermi gas, a Luttinger liquid behaves as the {\it Luttinger model}; 
a model describing spinless fermions in the continuum with linear dispersion relation
and short-range (non local) interaction.
The linear dispersion relation has the effect that infinitely many unphysical
fermions must be introduced to fill the ``Dirac sea'' of 
states with negative energy. This makes the model a bit unrealistic and
of no direct applicability to solid state physics but ,on the other hand, 
the choice of a linear dispersion relation has the effect that, contrary
to all other models of interacting fermions, the Luttinger model 
correlations can be explicitly computed, see [ML]. 
The popularity of the Luttinger liquid notion
increased greatly after the Anderson proposal [A] that the 
high-$T_c$ superconductors are, in their normal phase,
Luttinger liquids; this proposal was based 
on the conjecture that the Hubbard model 
in one or two dimensions have a somewhat similar behavior,  
and in particular that they {\it both} show Luttinger liquid behavior at least
for some range of the parameters. Up to now there is no 
agreement even at an heuristic level on
theoretical evidence of Luttinger liquid behavior in the $d=2$
Hubbard model. On the other hand the Anderson 
proposal stimulated a number of studied about
the Luttinger liquid behavior in $d=1$, as a natural prerequisite
to understand the same question in $d=2$. 

Numerical simulations of the correlation functions gave evidence 
[OS] that the $d=1$ Hubbard model is indeed a Luttinger liquid; subsequent
analytic (but heuristic) results by [PS] and [FK]  confirmed this result
{\it for large $U$}, finding also that the correlations
verify an important property, the 
{\it spin-charge separation}.
{\it For small $U$} the evidence for Luttinger liquid behavior 
is based on the two following facts:

1)the $d=1$ Hubbard model should be equivalent, 
as far as low energy property are considered, 
to a generalization of the Luttinger model
to {\it spinning} fermions, the so called {\it g-ology} model.

2)Contrary to the Luttinger model,
even the g-ology model is {\it not solvable}. However
Solyom [S] by Renormalization
Group (RG) analysis truncated at two loops 
showed that the $g-ology$ model scales iterating the RG
toward the {\it Mattis model},
a model which is indeed exactly solvable and it has Luttinger liquid behavior.

Given the above two facts, the formulas 
for the correlations of the $d=1$ Hubbard model are 
usually
approximated with the formulas for the correlations of the Mattis model,
see for instance [S]; this is however quite unsatisfactory
for a number of reasons:

1)this approximation means that the effects of the 
lattice and the corresponding Umklapp scattering terms
in the Hubbard model are completely neglected. 
There are however strong indications that this approximation
gives completely wrong predictions at least
for properties like the thermal or electric
conductivity [RA].

2)The conclusion of [S] that the Hubbard model scales iterating the RG 
toward the Mattis model is based on a number of peculiar cancellations in the
perturbative expansion, checked up to two loops.
Of course, without an argument stating that the cancellations
are present at {\it any order}, this conclusion is only perturbative
and not very solid; if at higher orders the cancellations were
not present, the effective
coupling constants could increase without limit making the analysis
meaningless. 

As a conclusion, the enormous number of results on the $d=1$
Hubbard model can be roughly divided in two main classes.
A first class are the exact results, based on the Bethe ansatz approach.
They are (almost) rigorous but they give essentially no informations
on the behavior of the correlations. Moreover
they are not very robust, as they rely on 
delicate integrability properties of the Hubbard model
and cannot be used
to face with
apparently harmless modifications of the Hubbard model
(for instance considering nonlocal but short ranged interactions).
The second class of results is obtained by a combination
of techniques (numerical simulations, bosonization, Renormalization Group)
and indeed informations on the correlations are found, but the results are 
not rigorous.

\* \sub(1.1a) {\it The Luttinger liquid construction}

In a series of paper 
[BM1], [BM2], [BM3], [BM4] a general proof of Luttinger
liquid behavior for {\it spinless} interacting fermions
(without any use of exact solutions) has been completed. 
The conclusion is that interacting {\it spinless} fermions
are {\it generically} Luttinger liquids (independently
from the dispersion relation, the presence of a lattice,
the sign of the interaction etc).
A perturbation theory for the correlation functions
is constructed, and such expansion is not in the strength of the interaction
but in terms of a set of parameters called {\it running coupling constants},
describing the effective interaction at a certain momentum
scale; the expansion is proved to be convergent (and analytic)
if the running coupling constants are small, see 
[BM1], as a consequence of suitable determinant bounds
for the fermionic truncated expectations. On the other hand
the property that the running coupling constants remain in the convergence radius
of the expansion is not trivial at all and is due to remarkable {\it cancellations}
at any order in the expansion. More exactly,
the running
coupling constants verify a set of recursive equations, 
whose l.h.s. is called {\it Beta function}, and
their boundedness is consequence of dramatic cancellations
happening at any order in the Beta function.
In order to prove such cancellations one decomposes the Beta function in the sum of  
two terms; one, called {\it dominant part},
which is common to all spinless $d=1$ Fermi system, and
the second part which depends on the specific model one is considering
and which gives a bounded flow {\it once one has proved that
the dominant part is asymptotically vanishing}. The problem
is then reduced to the vanishing of the dominant part
of the Beta function,which coincides with the complete
Beta function of a model, called {\it reference model},
describing interacting spinless fermions with an ultraviolet
and infrared cutoff. 
The proof of vanishing of the 
Beta function is the reduced to the proof of suitable (highly non trivial) 
Ward Identities between the correlation functions of the reference model
[BM3], for any value of the infrared cutoff.
The problem of implementing Ward identities in a model
with cutoffs (and in a Renormalization Group scheme) 
is a well known problem in Quantum field theory
or condensed matter physics. In 
[BM2], [BM4] a solution for this problem was given
by finding suitable
{\it Correction Identities} relating the corrections to the Ward
Identities due to cutoffs to the correlations themselves.
By combining
the Ward and the Correction Identities the vanishing of the reference model
Beta function is proved and the rigorous construction of the correlation
functions for spinless Luttinger liquid is then completed.

Aim of this paper is the extension of the Luttinger liquid construction
to the $d=1$ Hubbard model; such extension is not straightforward as
the Luttinger liquid is {\it not} the generic state
of {\it spinning} interacting fermions. The conditions of 
repulsive interaction $U>0$ and not half filling must be imposed;
technically this is reflected 
from the fact
that the expansions we find cannot be analytic
in a circle around $U=0$, as it would be in the spinless case, 
and $U$ must be chosen 
smaller and smaller as we are closer and closer to half-filling. 
We will define an expansion
for the correlations in terms of running coupling constants, but the presence
of the spin increases greatly their number; crucial symmetry 
considerations (based on the $SU(2)$ spin invariance of the Hubbard
model) and geometrical constraints reduce 
the number of the effective interactions (which is only one in the spinless case)
to 
three (in the non half filled band case)
or four (in the half filled band case). Again the 
running coupling constants verify a recursive relation,
whose r.h.s. is called Beta function, and the expansion
is meaningful only if the running coupling constant are small
at any momentum scale. One can decompose the Beta function
in a dominant part and a rest; it turns out however that the dominant part is not 
vanishing. 
Calling the three (in the not half filled case) effective interactions $g_1,g_2,g_4$,it turns out that,
truncating the beta function at the second order,
$g_1$ tends to
vanish (if $U>0$) while $g_2,g_4$
remains close to their initial value. In order to 
prove that such a result is valid non perturbatively,
that is including all orders contributions,
one has to prove a property which we will call
{\it partial vanishing} of the Beta function. 
Such property is derived
by a suitable {\it reference model}, which verifies
formally (if cutoffs are neglected) proper gauge symmetries.
Quite surprisingly, the cancellations on
Beta function of the Hubbard model, which verifies
an SU(2) spin symmetry, will be obtained by a reference model 
{\it not $SU(2)$ invariant}. We derive
suitable Ward and correction identities for the reference model, and by them
the partial vanishing of the Hubbard Beta function is proved.
Hence the running coupling constants are small
if $U$is small enough and the rigorous construction of the correlation
functions for Hubbard model is completed. 
The analysis can be easily
generalized to include the presence of a magnetic field or non local interactions.

\* \sub(1.2) {\it The Hubbard model}

The Hubbard model Hamiltonian is given by
$$H=-{t}\sum_{x\in\L}\sum_{\s=\pm}
(a^+_{x,\s}a^-_{x+1,\s}+a^+_{x+1,\s}a^-_{x,\s})+
U \sum_{x\in\L} a^+_{x,+}a^-_{x,+} a^+_{x,-}a^-_{x,-}-\m \sum_{x\in\L}\sum_{\s=\pm}
a^+_{x,\s}a^-_{x,\s}\Eq(1)$$
where $\L$ is an interval of $L$ points on the one dimensional lattice of step
$1$, which will be chosen equal to $(-[L/2],[(L-1)]/2)$ and $a^\pm_{x,\s}$
is a set of fermionic creation or annihilation operators 
with spin $\s=\pm$ satisfying
periodic boundary conditions; $t=1/2$ is the 
hopping parameter, $U>0$ the coupling and $\m$ the chemical potential.
The Hamiltonian verifyes an $SU(2)$ spin symmetry.

A generalization of the above hamiltonian including the effect
of a short-ranged (instead of a nearest-neighbor) interaction,
and the presence of a magnetic field, is the following 
$$H=-{1\over 2}\sum_{x,\s}(a^+_{x,\s}a^-_{x+1,\s}+a^+_{x+1,\s}a^-_{x,\s})+
U \sum_{x,y} v(x-y)a^+_{x,+}a^-_{x,+}a^+_{y,-}a^-_{y,-}-\m \sum_{x,\s}
a^+_{x,\s}a^-_{x,\s}\Eq(2)$$
$$+h(a^+_{x,+}a^-_{x,+}-a^+_{x,-}a^-_{x,-})$$
When the interaction is given by $U v(x-y)=U \d_{x,y}+ V
\d_{x+1,y}$ the model is known as the {\it $U-V$ model}.

We consider the operators 
$a^\pm_{\xx,\s}=e^{H x_0} a^\pm_{\xx,\s}e^{-H x_0}$,
$\xx=(x,x_0)$ and $-\b/2\le x_0\le\b/2$ for some $\b>0$; 
on $x_0$, which we call the time variable, periodic boundary
conditions are imposed.

Many physical properties of the fermionic system at inverse
temperature $\b$ can be expressed in terms of the {\sl Schwinger
functions}, that is the truncated expectations in the Grand Canonical
Ensemble of the time order product of the field $ a_{\xx}^{\pm}$ at
different space-time points. If 
$$<X>_{L,\b}={Tr e^{-\b H} X\over Tr e^{-\b H}}\Eq(Def)$$
the Schwinger functions are defined as,if $\e=\pm$
$$S_{L,\b}(\xx_1,\e_1,\s_1;...;\xx_1,\e_1,\s_1)=<a^{\e_1}_{\xx_1,\s_1}...a^{\e_1}_{\xx_n,\s_n}>_{L,\b}
\Eq(sf)$$
We will denote by $S(\xx_1,...)$ the $\lim_{L,\b\to\io}$ of \equ(sf).
An important role is played by the two point Schwinger function
$$S_{L,\b}(\xx,+,\s;\yy,-,\s)=S_{L,\b}(\xx,\yy)\Eq(sf11)$$
Denoting by $\hat S_{L,\b}(k,x_0)$ the Fourier transform of
$S_{L,\b}(\xx)$ with respect to the $x$ variable, $n_k\equiv \hat S_{L,\b}(k,0^-)$
is the {\it occupation number}, the average number of particles with
momentum $k$. Another important physical quantity
is the density-density 
{\it correlation function} 
$$\O_{\xx,\yy}=<\r_\xx \r_\yy>-<\r_\xx><\r_\yy>\Eq(ss)$$

\* \sub(1.2bb) {\it The non interacting $U=0$ case.}

The two point Schwinger function in the non interacting case is given by, in the
limit $L,\b\to\io$
$$S_0(\xx,\yy)=\int_{-\io}^\io dk_0 \int_{-\pi}^\pi dk {e^{-i\kk(\xx-\yy)}\over -i k_0+\m-\cos k}\Eq(1.2a)$$
where $\kk=(k,k_0)$.
It is easy to check that one can write, if $\m=\cos p_F^0$ and $v_0=\sin p_F^0$
$$S_0(\xx,\yy)=\sum_{\o=\pm} {e^{i \o p_F^0(x-y)}\over v_0 x_0+i \o x}+\bar g(\xx,\yy)\Eq(1.2aa)$$
with $|\bar g(\xx,\yy)|\le {C\over 1+|\xx-\yy|^{1+\th}}$, $\th$ a positive constant;
that is, the two point Schwinger function decays as $O(|\xx-\yy|^{-1})$
oscillating with period ${\pi\over p_F^0}$. Important 
physical properties are:

1)the occupation number
is given by $n_k=\chi(|\kk|\le p_F^0)$, that is is discontinuous.

2) The bidimensional Fourier transform of density correlation
has singularities at $(\pm 2p_F^0,0)$ and $(0,0)$; in 
$(\pm 2p_F^0,0)$ it has a logarithmic singularity while in $(0,0)$
the Fourier transform is bounded. 

3) The one dimensional Fourier
transform at $x_0=0$ of the density  correlation
is continuous, while its first derivative in $k$
has a first order discontinuity in 
$k=0, \pm 2 p_F^0$.

\* \sub(1.2vv) {\it Main result}

Our result can be informally stated in the following way
\*
{\it In the not half filled band case and in the weak 
coupling regime, the (repulsive) Hubbard model \equ(1) is
a Luttinger liquid.}
\*
A more formal statement is 
the following theorem.
\*
\sub(1.3) {\cs Theorem 1.} {\it Consider the hamiltonian \equ(1)
with $-1<\m<1$ and $\m\not=0$ (not filled or half filled band case); 
there exists an $\e>0$ such that, for 
$0<U<\e$ 

\0 a)the two point Schwinger function \equ(sf11) is given by, in the limit $L,\b\to\io$
$$S(\xx,\yy)=\sum_{\o=\pm} {e^{i \o p_F (x-y)}\over v (x_0-y_0)+i \o (x-y)}
{1+A_\o(\xx,\yy)\over |\xx-\yy|^\h}+\bar S(\xx,\yy)\Eq(1.3aa)$$
with 
$$\h=a U^2+U^2 f_0(U)\quad p_F=\cos^{-1}\m+f_1(U)\quad v=v_0+f_2(U) \Eq(prop)$$
where $a>0$, $|f_0(U)|, |f_1(U)|, |f_2(U)|\le C U$ and
$$|\bar\partial_x^{n_1} \partial_{x_0}^{n_0} A(\xx,\yy)|\le 
C U {1\over |\xx-\yy|^{n_0+n_1}}\quad |\bar S(\xx,\yy)|\le 
{C\over 1+|\xx-\yy|^{1+\th}}\Eq(prop1)$$
for suitable positive
constants $C,\th$, if $\bar\partial$ denotes the discrete derivative. Moreover the occupation number
$n_k$ is continuous at $k=\pm p_F$ but its first derivative diverges at $k=\pm p_F$ as 
$|k-(\pm p_F)|^{-1+\h}$.

\0 b) The density-density correlation function \equ(ss) can be written as
$$\O_{\xx,{\bf 0}}=\cos(2 p_F x) \O^{a}(\xx)
+\O^{b}(\xx)+\O^{c}(\xx)\;,\Eq(1.12)$$
with $\O^{i}(\xx)$, $i=a,b,c$, continuous bounded functions,
which are infinitely times differentiable as functions of $x_0$, if
$i=a,b$. Moreover
$$\eqalign{
\O^{a}(\xx) &= {1+A_1(\xx)\over 2\p^2[x^2+(v x_0)^2]^{1+\h_1}}\;,\cr
\O^{b}(\xx) &= {1\over 2\p^2[x^2+(v x_0)^2]}\Big\{ {x_0^2-(x/v_0)^2
\over x^2+(v_0 x_0)^2} + A_2(\xx)\Big\}\;,\cr}\Eq(1.19)$$
$$|A_i(\xx)| \le C U\quad\quad |\O^{o,c}(\xx)|\le {C\over 1+|\xx-\yy|^{2+\th}}
\;,\Eq(1.20)$$
for some constant $C$, where
$\h_1=-b U+U f_4(U)$ with $b>0$ and $|f_4(U)|\le C U$. Finally for $\a=a,b$ 
$$|\bar\partial_x^{n_1} \partial_{x_0}^{n_0} \O^{o,\a}(\xx)|\le
{C\over |\xx-\yy|^{2+n_0+n_1}}\Eq(hhjj)$$

\0 c) Let $\hat\O(\kk)$, $\kk=(k,k_0)\in [-\p,\p]\times \RRR^1$, the Fourier
transform of $\O(\xx)$. For any fixed $\kk$ with $\kk\not=(0,0), (\pm 2
p_F,0)$, $\hat\O(\kk)$ is uniformly bounded; moreover 
$$|\hat\O(\kk)|\le c_2[1+ U \log |\kk|^{-1}]\Eq(ffx)$$ 
near
$\kk=(0,0)$, and, at $\kk=(\pm 2p_F,0)$, diverges as 
$$|\kk-(\pm 2p_F,0)|^{2\h_1}/|\h_1|\Eq(kkn1)$$
Let $G(x)=\O(x,0)$ and $\hat G(k)$ its Fourier transform. 
Then $\dpr_k\hat G(k)$ has a first order discontinuity at $k=0$,
with a jump equal to $1+O(U)$, and, at $k=\pm 2p_F$, 
diverges as $|k-(\pm 2p_F)|^{2\h_1}$.
}
\*
{\bf Remarks}
\*
a)A naive estimate of $\e$ in the above Theorem is 
$\e=O(|\m|^{\a})$ 
for some constant $\a$,
for $\m$ close to $0$; that is $U$ must be taken smaller and smaller
as we are closer and closer to the half filled band case.
\*
b)A first effect of the interaction is that
the Fermi momentum $p_F$ is modified by the interaction by $O(U)$ terms.
\*
c)More dramatic is the effect of the interaction
on the long distance
asymptotic behavior of the physical observables;
it turns out that the two point
Schwinger function decays {\it faster} in presence of the interaction,
while the correlation function decays {\it slower}. The large
distance decay is power law with anomalous critical indexes depending
non trivially by the coupling $U$.
\*
d)As a consequence the occupation number $n_k$, which in the non interacting 
case have a discontinuity at $k=\pm p_F$, 
has no discontinuity in presence of the interaction; 
this proves that the $d=1$ Hubbard model is
a {\it Luttinger liquid} in the sense of [H]. The lack of discontinuity
in the occupation number can be physically interpreted 
saying that fermionic quasiparticles are not present.

The interaction changes the log-singularity at $\kk=(\pm 2p_F,0)$
of the bidimensional Fourier 
transform of the density correlation in 
a power law singularity, with a nonuniversal critical index $O(U)$.
This enhancement of the singularity 
is considered a signal of the tendency of the system
to develop density wave excitations with period $\pi/p_F$, generically incommensurate
with the lattice. On the other hand 
the singularity in $\kk=(0,0)$ is much weaker, that is at most logarithmic.

In the same way, the interaction leaves invariant the singularity
of the first derivative of the one dimensional
Fourier transform of the correlations in $k=0$ (a first order discontinuity)
while the singularity in $k=\pm 2 p_F$
is changed by the interaction from a discontinuity 
to a power law singularity.
\*
e)The two points Schwinger function and the density correlation
can be written as sum of two terms; one which is 
very similar to corresponding quantities in the Luttinger model, 
and in which the dependence from $p_F$ is quite simple
(they can be written as oscillating terms times terms which are
free of oscillations, in the sense that each derivative increases
the decay by a unit, see \equ(prop1),\equ(hhjj)) and another (non Luttinger like)
in which the dependence 
on $p_F$ and the lattice steps is very complicate; this last term
decays faster than the Luttinger like terms 
but the derivatives do not increase the decay
for the presence of oscillating terms. The non Luttinger like terms
have Fourier transform which is bounded;
however sufficiently high derivatives of the Fourier transform 
can be singular for values different respect to $k=0,\pm p_F,\pm 2 p_F$
(such singularities were indeed observed in numerical simulations,
see [OS]).
\*
d) Our results provide a proof of Luttinger liquid behavior,
but they are still not enough accurate to prove an important property called {\it
spin-charge separation}, which is believed true for the Hubbard model; namely
that the asymptotic behavior of the two point Schwinger function is
$(x_0+i v_c x)^{-{1\over 2}-\h_c}(x_0+i v_s x)^{-{1\over 2}-\h_s}$,
with $v_c-v_s=O(U)$ and $\h_c,\h_s=O(U)$; \equ(1.3aa),\equ(prop1) is compatible
with such behavior but is not enough accurate to prove it. Another
property which could be probably proved by an extension of our techniques
is the Borel summability of our critical indexes as a function of $U$.
\* 
e)Finally we could consider a short range instead of local 
potential, that is \equ(2) with $h=0$. In such a case
the condition $U>0$ is replaced by the condition
$U\hat v(2p_F)+F(U)>0$, where $F(U)$ is a suitable $O(U^2)$
function. Note that the linear term is vanishing for 
sufficiently long range interactions such that $\hat v(2 p_F)=0$.

\* \sub(1.aa2) {\it The Hubbard model in a magnetic field}

Let us consider the Hamiltonian \equ(2) with $h\not=0$; the
presence of a magnetic field destroys the spin rotation invariance.
Moreover it turns out that that one can consider also attractive
interactions, if the interaction is smaller than the magnetic field. Calling $S_{\s,L,\b}(\xx,\yy)
=<\psi^-_{\xx,\s}\psi^+_{\yy,\s}>_{L,\b}$ we prove the following result.
\*
\sub(1.3a) {\cs Theorem 2.} {\it Consider the hamiltonian \equ(2)
with $-1<\m<1$ and $0\le h\le h_0$ for a suitable constant $h_0$ ; assume also that $\cos^{-1}(\m+h)
+\cos^{-1}(\m-h)\not=\pi$.
There exists positive constants 
$\e_1,\e_2$ (depending on $\m$ and $h$,and $\e_2$ vanishing as $h\to0$) 
such that,
if $-\e_2\le U\le \e_1$ the two point Schwinger function is given by, in the limit $L,\b\to\io$
$$S_\s(\xx,\yy)=\sum_{\o} {e^{i \o p_F^\s (x-y)}\over v (x_0-y_0)+i \o 
(x-y)}{1+A_\o(\xx,\yy)\over |\xx-\yy|^\h}+\bar S(\xx,\yy)\Eq(1.3aa1)$$
with 
$$\h=a U^2+O(U^3)\quad p_F^\s=\cos^{-1}(\m+\sign(\s)h)+
O(U)\quad v=\sin (\cos^{-1}(\m+\sign(\s)h))+O(U)\Eq(prop1)$$
where $a>0$ and
$$|\partial^{n_0}\bar\partial^{n_1} 
A(\xx,\yy)|\le C U |\xx-\yy|^{-n_0-n_1}\quad |\bar S(\xx,\yy)|\le {C\over 1+|\xx-\yy|^{1+\th}}\Eq(prop11)$$
for suitable positive
constants $C,\th$. } 
\*
The other statements 
in the previous Theorem can be repeated with some obvious modifications.
The above result says that the Hubbard model  
is still a Luttinger liquid even in presence of a magnetic field;
this happens even in the attractive case,if the interaction is smaller 
than the magnetic field.
\*
\sub(1.aa2x) {\it Contents}

In \S 2 and \S 3 we write the Hubbard model \equ(1) partition function
as a Grassmann integral, and we define a multiscale integration procedure;
we get an expansion in terms of running coupling constants, whose 
regularity properties are stated in Theorem 3. In \S 4 we study the flow
of the running coupling constants and in \S 5 and \S 6 we derive
the cancellations of the Hubbard model Beta function by Ward identities
and Correction identities of a suitable reference model. Finally such results are applied
in \S 7 to the computation of the Schwinger functions and the correlations and in
\S 8 the presence of the magnetic field is included. 
We rely on many technical results
already obtained in  [BM1-4] (the presence of spin 
has a small effect on the proof of convergence, for instance)
and we focus mainly on the difference with respect to the spinless case.

\vskip.5cm
\section(2, The ultraviolet integration)
\vskip.5cm
We assume $\m\in\O\cap(-1,1)$, where $\O^c=\{\m:|\m|,
|\m\pm 1|\le \bar\m$, where $\bar\m>0$ is a fixed constant.
We call ${\cal D}_{L,\b}\={\cal D}_L
\times {\cal D}_\b$, with ${\cal D}_L\=\{k={2\pi n/L}, n\in \zzzz, 
-[L/2]\le n \le [(L-1)/2]\}$ and
${\cal D}_\b\=\{k_0=2(n+1/2)\pi/\b, n\in
Z, -M\le n \le M-1\}$;  moreover we define
$$\tilde t=1-\d\quad \tilde t\cos p_F=\m-\n\Eq(bb)$$
with $\d, \n$ suitable counterterms to be fixed
properly in the following. 
We introduce a finite set of Grassmanian
variables $\{\hat \psi^\pm_\kk\}$, one for each $\kk\in\DD_{L,\b}$, and a
linear functional $P(d\psi)$ on the generated Grassmannian algebra, such that
$$\int P(d\psi) \hat \psi^-_{\kk_1,\s_1}
\hat \psi^+_{\kk_2,\s_2} = L\b \d_{\kk_1,\kk_2}\d_{\s_1,\s_2}
\hat g(\kk_1)\;,\quad \hat g(\kk)= {1\over -ik_0+\tilde t \cos p_F-\tilde t \cos k}
\; .\Eq(2.8)$$
We will call $\hat g(\kk)$ the {\it propagator} of the field.

We define also {\sl Grassmanian field} $\psi^\pm_\xx$ is defined by
$$\psi_{\xx,\s}^{\pm}= {1\over L\b} \sum_{\kk\in {\cal D}_{L,\b}}\hat \psi_{\kk,\s}^{\pm}
e^{\pm i\kk\cdot\xx}\Eq(chic)$$
such that
$${1\over L\b} \sum_{\kk\in {\cal D}_{L,\b}} \, e^{-i\kk\cdot(\xx-\yy)} \,
\hat g(\kk) = \int P(d\psi) \psi^-_\xx  \psi^+_\yy
\= g^{L,\b}(\xx;\yy) \; ,\Eq(2.9)$$
It is well known that the partition function  $Z=<e^{-\b H}>_{L,\b}$
can be rewritten as the limit $M\to\io$
of the Grassmann integral 
$$\int P(d\psi) e^{-\VV}\Eq(h1)$$
where $P(d\psi)$ is the Grassmann integration with propagator \equ(2.9)
and
$$\VV=U \int_{-\b/2}^{\b/2} dx_0 \sum_x 
\psi^+_{\xx,+}\psi^-_{\xx,+}\psi^+_{\xx,-}\psi^-_{\xx,-}+\nu 
\int_{-\b/2}^{\b/2} d x_0 \sum_{x,\s}
\psi^+_{\xx,\s}\psi^-_{\xx,\s}+\d 
\int_{-\b/2}^{\b/2} d x_0 \sum_{x,y,\s}t_{x,y} 
\psi^+_{\xx,\s}\psi^-_{\xx,\s}\Eq(v)$$
where $t_{x,y}
={1\over 2}\d_{y,x+1}+{1\over 2}\d_{x,y+1}$ and
$P(d\psi)$ has propagator given by 
$g(\xx,\yy)$ \equ(2.9).
Let $T^1$ be the one dimensional torus, $||k-k'||_{T^1}$ the usual
distance between $k$ and $k'$ in $T^1$ and $||k||=||k-0||$.
We introduce a {\sl scaling parameter} $\g>1$ and a positive function
$\c(\kk') \in C^{\io}(T^1\times R)$, $\kk'=(k',k_0)$, such that
$$ \c(\kk') = \c(-\kk') = \cases{
1 & if $|\kk'| <t_0 \= a_0 v_0 /\g \;,$ \cr
0 & if $|\kk'| >a_0 v_0\; ,$\cr}\Eq(2.30)$$
where
$$|\kk'|=\sqrt{k_0^2+(v_0 ||k'||_{T^1})^2}\;,\Eq(2.31)$$
$$a_0 =\min \{p_F/2, (\p-p_F)/2 \}\;,\Eq(2.32)$$

The definition \equ(2.30) is such that the supports of $\c(k-p_F,k_0)$ and
$\c(k+p_F,k_0)$ are disjoint and the $C^\io$ function on $T^1\times R$
$$\hat f_{u.v.}(\kk) \= 1- \c(k-p_F,k_0) - \c(k+p_F,k_0) \Eq(2.35)$$
is equal  to $0$, if $[v_0||(|k|-p_F)||_{T^1}]^2 +k_0^2<t_0^2$.
We define
$$g^{L,\b}(\xx;\yy)=g^{u.v.}(\xx,\yy)+g^{i.r.}(\xx,\yy)\Eq(dec)$$
with 
$$g^{u.v.}(\xx,\yy)={1\over L\b}
\sum_{\kk\in\DD_{L,\b}}e^{-i\kk(\xx-\yy)}{\hat f_{u.v.}(\kk)\over -i k_0-\tilde t\cos k+\tilde t\cos p_F}$$
$$g^{i.r.}(\xx,\yy)={1\over L\b}
\sum_{\kk\in\DD_{L,\b}}e^{-i\kk(\xx-\yy)}{\prod_{\o=\pm 1}\c(k-\o p_F,k_0)\over -i k_0-\tilde t
\cos k+\tilde t\cos p_F} \Eq(1.2b)$$
From the integration over $\psi^{(u.v.)}$ we get
$$e^{-L\b E_{L,\b}} = e^{-L\b (\tilde E_1+t_1)} \int P(d\psi^{(i.r.)}) \,
e^{-\VV^{(0)}(\psi^{(i.r.)})}
\;,\quad \VV^{(0)}(0)=0\;,\Eq(2.59)$$
$$e^{-\VV^{(0)}(\psi^{(i.r.)})-L\b \tilde E_1}= \int
P(d\psi^{(u.v.)}) e^{-\VV(\psi^{(i.r.)}+\psi^{(u.v)}})\;.\Eq(2.60)$$
We will call $\psi^{(i.r.)}=\psi^{(\le 0)}$;
$\VV^{(0)}(\psi^{(\le 0)})$ can be written in the form
$$\eqalign{
&\VV^{(0)}(\psi^{(\le 0)}) = \sum_{n=1}^\io {1\over (L\b)^{2n}}
\sum_{\ss} \sum_{\kk_1,...,\kk_{2n}}
\prod_{i=1}^{2n} \hat\psi^{(\le 0)\e_i}_{\kk_i,\s_i}\;\cdot\cr
&\cdot\;{\hat W}_{2n,\ss,\oo}^{(0)}(\kk_1,...,\kk_{2n-1})\;
\d(\sum_{i=1}^{2n}\e_i\kk_i)\;,\cr}\Eq(2.61)$$
where $\ss=(\s_1,\ldots,\s_{2n})$, $\oo=(\o_1,\ldots,\o_{2n})$ and
we used the notation
$$\d(\kk)=\d(k)\d(k_0)\;,\quad\d(k)=L \sum_{n\in\zzzz} \d_{k,2\p n}\;,\quad
\d(k_0)=\b\d_{k_0,0}\;.\Eq(2.62)$$
We prove in the Appendix that
$$|{\hat W}_{2n,\ss,\oo}^{(0)}(\kk_1,...,\kk_{2n-1})|
\le L\b C \max(U,|\n|)^{\max(1,n/2)}\Eq(2.62a)$$
The $SU(2)$ spin invariance implies that the quartic terms have
the following form
$${1\over (L\b)^4}\sum_{\kk_1,..,\kk_4} W^{(0)}_{4,\oo}(\kk_1,...,\kk_4)\sum_{\s}\d(\sum_i \e_i\kk_i)
[\psi^+_{\kk_1,\s}\psi^-_{\kk_2,\s}
\psi^+_{\kk_3,\s}\psi^-_{\kk_4,\s}+
\psi^+_{\kk_1,\s}\psi^-_{\kk_2,\s}
\psi^+_{\kk_3,-\s}\psi^-_{\kk_4,-\s}]\Eq(2.62b)$$
where $W^{(0)}_{4,\oo}$
is spin independent. 
\vskip.5cm
\section(3, The infrared integration)
\vskip.5cm
\* \sub(3.1) {\it Quasiparticles.}

We define also, for any integer $h\le 0$,
$$f_h(\kk')= \c(\g^{-h}\kk')-\c(\g^{-h+1}\kk')\; ;\Eq(2.36)$$
we have, for any $\bh<0$,
$$\c(\kk') = \sum_{h=\bh+1}^0 f_h(\kk') +\c(\g^{-\bh}\kk')\; .
\Eq(2.37)$$
Note that, if $h\le 0$, $f_h(\kk') = 0$ for $|\kk'|
<t_0\g^{h-1}$ or $|\kk'| >t_0 \g^{h+1}$, and $f_h(\kk')=
1$, if $|\kk'| =t_0\g^h$.
We finally define, for any $h\le 0$:
$$ \hat f_h(\kk) = f_h(k-p_F,k_0) +f_h(k+p_F,k_0)\;;\Eq(2.39)$$
This definition implies that, if $h\le 0$, the support of
$\hat f_h(\kk)$ is the union of two disjoint sets, $A_h^+$ and $A_h^-$. In
$A_h^+$, $k$ is strictly positive and $||k-p_F||_{T^1}\le a_0\g^h \le a_0$,
while, in $A_h^-$, $k$ is strictly negative and $||k+p_F||_{T^1}\le a_0\g^h$.
 The label $h$ is called the {\sl scale} or {\sl frequency} label.
Note that, if $\kk\in {\cal D}_{L,\b}$, then $|\kk\pm (p_F,0)|\ge
\sqrt{(\p\b^{-1})^2+
(v_0\p L^{-1})^2}$, by the definition of ${\cal D}_{L,\b}$.
Therefore
$$\hat f_h(\kk)=0\quad \forall h< h_{L,\b} =\min \{h:t_0\g^{h+1} >
\sqrt{(\p\b^{-1})^2+(v_0\p L^{-1})^2} \}\;,\Eq(2.40)$$
and, if $\kk\in {\cal D}_{L,\b}$, the definitions \equ(2.35) and
\equ(2.39), together with the identity \equ(2.37), imply that
$$1=\sum_{h=h_{L,\b} }^0 \hat f_h(\kk)+\hat f_{u.v.}(\kk) \; .\Eq(2.41)$$
We now introduce, for any $h\le 0$, a set of Grassmann variables $\psi^\pm_{\kk',\o}$
such that
$$\int P(d\psi^{(h)}) \psi^{-(h)}_{\kk'_1,\o,\s} 
\psi^{+(h)}_{\kk'_2,\o',\s'}=L\b \d_{\s,\s'}\d_{\o,\o'}\d_{\kk'_1,\kk'_2}g^{(h)}_\o (\kk'_1)
\;.\Eq(2.43aab)$$
where
$$g^{(h)}_\o (\kk')=
{f_h(k',k_0)\over -i k_0-\tilde t \cos (k'+\o p_F)+\tilde t\cos p_F}\Eq(ppl)$$
We introduce also the Grassmann variables
$$\psi_{\xx,\o,\s}^{\pm (h)}= {1\over L\b} 
\sum_{\kk'\in {\cal D}_{L,\b}}\hat \psi_{\kk',\o,\s}^{\pm (h)}
e^{\pm i\kk'\cdot\xx}
\Eq(chic1)$$
so that
$$\int P(d\psi^{(h)})
\psi^{-(h)}_{\xx,\o,\s} 
\psi^{+(h)}_{\yy,\o',\s'}=\d_{\s,\s'}\d_{\o,\o'}g^{(h)}_\o (\xx,\yy)
\;.\Eq(2.43aab)$$
where
$$g^{(h)}_\o (\xx,\yy)={1\over L\b}
\sum_{\kk'\in\DD_{L,\b}}e^{-i\kk'(\xx-\yy)}
g_{\o}^{(h)}(\kk')
\;. \Eq(1.2bnn)$$
It holds that 
$$\int P(d\psi^{(i.r.)})
\psi^{-(i.r.)}_{\xx,\s} 
\psi^{+(i.r.)}_{\yy,\s'}=
\d_{\s,\s'}\sum_{h=h_{L,\b}}^0\sum_{\o=\pm} e^{-i\o p_F(x-y)} 
g^{(h)}_\o (\xx,\yy)
\;.\Eq(2.43aalb)$$
The above identity implies that, if $F(\psi^{(i.r.)})$ 
is any analytic function of the variables $\psi^{(i.r)}$ 
$$\int P(d\psi^{(i.r.)})F(\psi^{(i.r.)})=\int \prod_{h=h_{L,\b}}^0 
\int P(d\psi^{(h)}) F(\sum_{h=h_{L,\b}}^0
\sum_{\o=\pm} e^{-i\o p_F x} \psi_{\xx,\s}^{(h)})
\Eq(decc)
$$
We define also
$$C_h^{-1}(\kk)=\sum_{k=-\io}^h \hat f_k(\kk)\Eq(kkms)$$
\* \sub(3.2) {\it Multiscale integration.}
The integration of the infrared part is done in an iterative way.
Assume that we have integrated the scales $0,-1,...,h+1$
and that we have found
$$\int P_{Z_h,C_h}(d\psi^{(\le h)}) \, e^{-\VV^{(h)}
(\sqrt{Z_h}\psi^{(\le h)})-L\b E_h}\;,\quad \VV^{(h)}(0)=0\;,\Eq(4.3a)$$
where  
$$P_{Z_h,C_h}(d\psi^{(\le h)})=\NN_h^{-1}\prod_{\kk'\in D} \prod_{\o=\pm 1}
d\psi^{+(\le h)}_{\kk',\o}d\psi^{-(\le h)}_{\kk',\o}$$
$$\exp -{1\over \b L}\sum_{\kk'\in D \atop C_h^{-1}(\kk)>0}
Z_h C_h(\kk) \psi_{\kk',\o,\s}^{+(\le h)} (-ik_0+\o v_0\sin k'+\cos p_F(\cos k'-1))
\psi_{\kk',\o,\s}^{-(\le h)}
\Eq(4.4a)$$
and 
$$\eqalign{
&\VV^{(h)}(\psi^{(\le h)}) = \sum_{n=1}^\io {1\over (L\b)^{2n}}
\sum_{\ss} \sum_{\kk_1',...,\kk'_{2n}}
\prod_{i=1}^{2n} \hat\psi^{(\le 0)\e_i}_{\kk'_i,\o_i,\s_i}\;\cdot\cr
&\cdot\;{\hat W}_{2n,\ss,\oo}^{(h)}(\kk'_1,...,\kk'_{2n-1})\;
\d(\sum_{i=1}^{2n}\e_i\kk'_i+\sum_{i=1}^{2n} \e_i\o_i p_F
)\;,\cr}\Eq(2.61a)$$
and in particular the quartic terms have the following form
$${1\over (L\b)^4}\sum_{\kk'_1,..,\kk'_4} \sum_{\oo} W^{(h)}_{4,\oo}(\kk'_1,...,\kk'_3)\sum_{\s}
\d(\sum_i\e_i \kk'_i+\sum_i\e_i\o_i p_F)$$
$$[\psi^+_{\kk_1,\o_1,\s}\psi^-_{\kk_2,\o_2,\s}
\psi^+_{\kk_3,\o_3,\s}\psi^-_{\kk_4,\o_4,\s}+
\psi^+_{\kk_1,\o_1,\s}\psi^-_{\kk_2,\o_2,\s}
\psi^+_{\kk_3,\o_3,-\s}\psi^-_{\kk_4,\o_4,-\s}]\Eq(qq)$$
Note that there exists a scale $\bar h$ such that, for $h\le\bar h$
are present in \equ(qq)
only the monomials verifying 
$$||\sum_{i=1}^4 \e_i \o_i p_F||_{T^1}=0\,.\Eq(cond)$$
In fact by the compact support properties of the propagators 
$||\sum_i\e_i \kk'_i||_{T^1}\le 4 a_0 v_0 \g^{h+1}$ and if \equ(cond)
is not satisfied $||\sum_{i=1}^4 \e_i \o_i p_F||_{T^1}\ge
C |p_F-{\pi\over 2}|$; hence $\bar h=O(\log |p_F-{\pi\over 2}|)$.
The condition $|\m|\ge \bar\m$
surely implies that $|p_F-{\pi\over 2}|>0$ (hence that $\bar h$ is finite if $\m\not=0$), 
for $U$ small enough (than $O(\bar\m)$), 
as $p_F=\cos^{-1}\m+O(U)$.

\* \sub(3.3) {\it The localization operator.}

We split the effective potential
$\VV^{(h)}$ as $\LL \VV^{(h)}+\RR \VV^{(h)}$, where
$\RR=1-\LL$ and $\LL$, the {\it localization operator}, is a linear operator
on functions of the form \equ(2.61a), defined in the following way by its action
on the kernels $\hat W_{2n,\oo}^{(h)}$.
\*
1) If $2n=4$ we define
$$\LL \hat W_{4,\ss,\oo}^{(h)}
(\kk'_1,\kk'_2,\kk'_3)=L^{-1}\d(\sum_{i=1}\e_i \o_i p_F)
\hat W_{4,\ss,\oo}^{(h)}(\bk++,\bk++,\bk++)\;,\Eq(1.18)$$
where
$\bk\h{\h'} = (\h\p L^{-1},\h'\p\b^{-1})$.
\*
2) If $2n=2$ (in this case there is a non zero contribution only
if $\o_1=\o_2$)
$$\LL \hat W_{2,\ss,\oo}^{(h)}(\kk')= \fra14 \sum_{\h,\h'=\pm 1}
\hat W_{2,\ss,\oo}^{(j)}(\bk\h{\h'})\left\{ 1+ \h {L\over \p} +
\h'{\b\over \p} k_0 \right\}\;,\Eq(1.19)$$
\*
3) In all the other cases
$$\LL \hat W_{2n,\ss,\oo}^{(h)}(\kk'_1,\ldots,\kk'_{2n-1})=0\;.\Eq(1.20)$$
\*
In the not half filled band case $p_F\not={\pi\over 2}$ 
the condition $\d(\sum_{i=1}\e_i \o_i p_F)\not=0$ is equivalent to
the condition $\sum_{i=1}^4
\e_i\o_i\not=0$.
Then the action of $\LL$ if $n=2$ is non trivial only if
$\sum_{i=1}^4 \e_i \o_i=0$ and there are only the following
possibilities for $\o_1,\o_2,\o_3,\o_4$:
$$(\o,\o,-\o,-\o);\quad
(\o,-\o,-\o,\o);\quad(\o,\o,\o,\o)\Eq(poss)$$
In the half filled band case $p_F={\pi\over 2}$ the action of $\LL$
is non trivial also if $\o_1=\o_3=-\o_2=-\o_4$.

We get 
$$\LL\VV^{(h)}(\psi)=\g^h n_h F_\n^{(h)}(\psi)+z_h F_z^{(h)}(\psi) 
+a_h F_a^{(h)}(\psi)+\g_{1,h} F_1^{(h)}(\psi)+\g_{2,h} F_2^{(h)}(\psi)
+\g_{4,h} F_4^{(h)}(\psi)\Eq(pef)$$
where
$$F_\n={1\over\b L}\sum_{\kk'}\sum_{\o,\s}\psi^+_{\kk',\o,\s}\psi^-_{\kk',\o,\s}$$
$$F_z={1\over\b L}\sum_{\kk'}(- i k_0)\sum_{\o,\s}\psi^+_{\kk',\o,\s}\psi^-_{\kk',\o,\s}$$
$$F_a={1\over\b L}\sum_{\kk'}[\o\sin p_F\sin k'+\cos p_F (\cos k'-1)]\sum_{\o,\s}\psi^+_{\kk',\o,\s}\psi^-_{\kk',\o,\s}$$
$$F_1={1\over(\b L)^4}\sum_{\kk'_1,\kk'_2,\kk'_3,\kk'_4}
\sum_{\o}\sum_{\s,\s'}\d(\sum_i\e_i\kk'_i)
\psi^+_{\kk'_1,\o,\s}\psi^-_{\kk'_2,-\o,\s}\psi^+_{\kk'_3,-\o,\s'}\psi^-_{\kk'_4,\o,\s'}$$
$$F_2={1\over(\b L)^4}\sum_{\kk'_1,\kk'_2,\kk'_3,\kk'_4}\sum_{\o,\s,\s'}
\d(\sum_i\e_i\kk'_i)
\psi^+_{\kk'_1,\o,\s}\psi^-_{\kk'_2,\o,\s}\psi^+_{\kk'_3,-\o,\s'}\psi^-_{\kk'_4,-\o,\s'}$$
$$F_4={1\over(\b L)^4}\sum_{\kk'_1,\kk'_2,\kk'_3,\kk'_4}\sum_{\o,\s}\d(\sum_i\e_i\kk'_i)
\psi^+_{\kk'_1,\o,\s}\psi^-_{\kk'_2,\o,\s}\psi^+_{\kk'_3,\o,\s'}\psi^-_{\kk'_4,\o,\s'}$$
Note that
$$\g_{2,h}=\hat W^{(h)}(\o p_F,\o p_F,-\o p_F,-\o p_F)
\quad \g_{1,h}=\hat W^{(h)}(\o p_F,-\o p_F,-\o p_F,\o p_F)$$
$$\quad \g_{4,h}=\hat W^{(h)}(\o p_F,\o p_F,\o p_F,\o p_F)\Eq(nn)$$
and in particular
$$\g_{4,0}=U \hat v(0)+O(U^2)\quad \g_{2,0}=U \hat v(0)+O(U^2)\quad
\g_{1,0}=U \hat v(2p_F)+O(U^2)$$
In the case of local interactions $v(p)=1$.
Note also that the spin symmetric part of $g^4$ is
vanishing by Pauli principle.

\vskip.5cm
\* \sub(3.4) {\it Renormalization.}

We write \equ(4.3a) as
$$\int P_{Z_h,C_h}(d\psi^{(\le h)}) \, e^{-\LL\VV^{(h)}
(\sqrt{Z_h}\psi^{(\le h)})-\RR\VV^{(h)}
(\sqrt{Z_h}\psi^{(\le h)}) -L\b E_h}\;,\Eq(4.3a)$$
and we include the quadratic part of  
$\LL\VV^{(h)}$ given by $z_h \int d\kk' \sum_{\o,\s}
\psi^+_{\kk',\o,\s}(-ik_0+\o\sin k'+\cos p_F(\cos k'-1))\psi^-_{\kk',\o,\s}$
in the free integration; 
we call 
$$\LL\bar\VV^h=\LL\VV^{(h)}-z_h \int d\kk' \sum_{\o,\s}
\psi^+_{\kk',\o,\s}(-ik_0+\o\sin k'+\cos p_F(\cos k'-1))\psi^-_{\kk',\o,\s}\Eq(99o)$$
so that we obtain
$$\int P_{\tilde Z_{h-1},C_h}(d\psi^{(\le h)}) \, e^{
-\LL\bar\VV^h(\sqrt{Z_h}\psi^{(\le h)})-
\RR\VV^{(h)}
(\sqrt{Z_h}\psi^{(\le h)}) -L\b E_h}\;,\Eq(4.3aa)$$
where
$$\tilde Z_{h-1}(\kk)\defin Z_h (1+z_h C_h^{-1}(\kk))\Eq(4.16a)$$ 
It is convenient to rescale the fields:
$${\widehat\VV}^{(h)}(\sqrt{Z_{h-1}}\psi^{(\le h)})\ \defin\ 
g_{1,_h} F_1(\sqrt{Z_{h-1}}\psi^{(\le h)})+
g_{2,_h} F_2(\sqrt{Z_{h-1}}\psi^{(\le h)}+
g_{4,_h} F_1(\sqrt{Z_{h-1}}\psi^{(\le h)}+$$
$$+\d_h F_a(\sqrt{Z_{h-1}}\psi^{(\le h)}+
\g^h\n_h F_\n(\sqrt{Z_{h-1}}\psi^{(\le h)})
+\RR\VV^{(h)}(\sqrt{Z_h}\psi^{(\le h)})
\;,\Eq(4.18)$$
where
$$\n_h={Z_h\over Z_{h-1}}n_h\quad \d_h={Z_h\over Z_{h-1}}[a_h-z_h]\quad g_{i,h}=[{Z_h\over Z_h-1}]^2 \g_{1,h}\Eq(ppl)$$
Finally the r.h.s. of \equ(4.3aa) can be rewritten as
$$e^{-L\b t_h}\int P_{Z_{h-1},C_{h-1}}(d\psi^{(\le h-1)}) 
\int P_{ Z_{h-1},\tilde f_{h}^{-1}}(d\psi^{(h)}) 
\, e^{-\hat\VV^{(h)}(\sqrt{Z_{h-1}}\psi^{(\le h)})}\;,\Eq(4.3aa)$$
where 
$$Z_{h-1}=Z_h(1+z_h)\quad
\tilde f_h(\kk')=f_h(\kk')[1+{z_h f_{h+1}(\kk')\over 1+ z_h f_h(\kk')}]\Eq(oop)$$
and
$$\int  P_{ Z_{h-1},\tilde f_{h}^{-1}}(d\psi^{(h)}) 
\psi^{-(h)}_{\xx,\o_1,\s_1} 
\psi^{+(h)}_{\yy,\o_2,\s_2}=
\d_{\s_1,\s_2}\d_{\o_1,\o_2}
{\tilde g^{(h)}_\o (\xx,\yy)\over Z_{h-1}}=$$
$${1\over Z_{h-1}}{1\over L\b}\sum_{\kk'\in D_{L,\b}}\tilde f_h(\kk){e^{-i\kk'(\xx-\yy)}\over -ik_0+\o v_0\sin k'+\cos p_F(\cos k'-1)}
\;.\Eq(2.43aabc)$$
We then integrate $\psi^{(h)}$
$$\int P_{Z_{h-1},\tilde f_{h}^{-1}}(d\psi^{(\le h)}) 
\, e^{-\hat\VV^{(h)}(\sqrt{Z_{h-1}}\psi^{(\le h)})}=e^{-\VV^{(h-1)}(\sqrt{Z_{h-1}}\psi^{(\le h)})}\Eq(hh)$$
and the procedure can be iterated.

Note that the quartic terms in $\LL\VV^h$ can be written in
coordinate representation in the following way
$$\sum_{\o,\s} \int d\xx  g_2^h [\psi^+_{\xx,\o,\s}\psi^-_{\xx,\o,\s}
\psi^+_{\xx,-\o,\s}\psi^-_{\xx,-\o,\s}+
\psi^+_{\xx,\o,\s}\psi^-_{\xx,\o,\s}
\psi^+_{\xx,-\o,-\s}\psi^-_{\xx,-\o,-\s}]$$
$$+g_1^h [\psi^+_{\xx,\o,\s}\psi^-_{\xx,-\o,\s}
\psi^+_{\xx,-\o,\s}\psi^-_{\xx,\o,\s}+
\psi^+_{\xx,\o,\s}\psi^-_{\xx,-\o,\s}
\psi^+_{\xx,-\o,-\s}\psi^-_{\xx,\o,-\s}]\Eq(bb1)$$
$$+g_4^h [
\psi^+_{\xx,\o,\s}\psi^-_{\xx_2,\o,\s}
\psi^+_{\xx,\o,-\s}\psi^-_{\xx,\o,-\s}]$$
where $\int d\xx=\int d x_0 \sum_x$.
Finally note that
the propagator is written as
$$\tilde g^h_\o(\xx-\yy)=g^h_{1,L}(\xx-\yy)+r^h_{1,L}(\xx-\yy)\Eq(dec)$$
where
$$g^h_{1,L}(\xx-\yy)=
{1\over\b L}\sum_\kk f_h(\kk) {e^{-i\kk(\xx-\yy)}\over -ik_0+\o k}$$
and for any positive integer $N$ 
$$|r_h(\xx-\yy)|\le C_N {\g^{2 h}\over 1+(\g^h|\xx-\yy|)^N}$$
It is easy to verify that $g_L^{(h)}$ verifies the same bound of $r_h$ with a $\g^h$ less.
We call
$v_k=(\n_k,\d_k,g_{1,k},g_{2,k}, g_{4,k})$, $k\le 0$
and $v_1=(\n,\d,U)$; moreover we call
$g_k=(g_{1,k},g_{2,k} g_{4,h})$ and $\m_k=(g_{2,k}, g_{4,k})$, $k\le 0$.
The above integration procedure generates a power series 
expansion for $W^{(h)}_{2n,\ss,\oo}(\xx_1,..,\xx_{2n})$
in \equ(2.61a) in terms of the running coupling constants $\vec v_k
$, $k=1,0,-1,-2,...,h$, which 
is indeed convergent if they are small enough. More exactly it holds
the following result. 
\vskip.5cm
\* 
\sub(3.3hh) It holds the following crucial result.
\*
{\cs Theorem 3.} {\it Assume that $\m\not =0,\pm 1$ and $\sup_{k\ge h}|\vec v_k|\le\e_h $;
assume also that, for some constant $c$, $\sup_{k\ge h} {Z_k\over Z_{k-1}}\le e^{c\e_h^2}$;
then there exists $\bar\e$ such that, for $\e_h\le\bar\e$
the functions $W^{(h)}_{2n,\ss,\oo}(\xx_1,...\xx_n)$ are analytic in the running coupling 
constants $(\vec v_k)_{k\ge h}$
and, for a suitable constant $C,\a$ 
$$\int d\xx_1..d\xx_n| W^{(h)}_{2n,\ss,\oo}
(\xx_1,...\xx_n)|\le (C \e_h\bar h^\a)^{\max(1,n/2)} L\b \g^{(2-n)h}\Eq(xx)$$
}
\*
{\it Sketch of the proof.} The proof is essentially
identical to the one of Theorem (3.12) of [BM1] about  
the spinless case.  The only important difference is that 
there exists a finite scale $\bar h=O(\log |p_F-{\pi\over 2}|)$ 
such that for $h\le\bar h$
there are no contributions to the effective potential $\bar\VV^h$
\equ(2.61a) with $n=2$ and a choice of $\o,\e$
such that \equ(cond) is not verified.
We can write the effective
potential $\VV^{(h)}(\sqrt{Z_h}\psi^{(\le h)})$, for $h\le 0$,
in terms of a {\it tree expansion}, similar to that described in [BM1].

\insertplot{300pt}{150pt}%
{\input treelut.txt}%
{treelut}{\eqg(1)}
\*\*
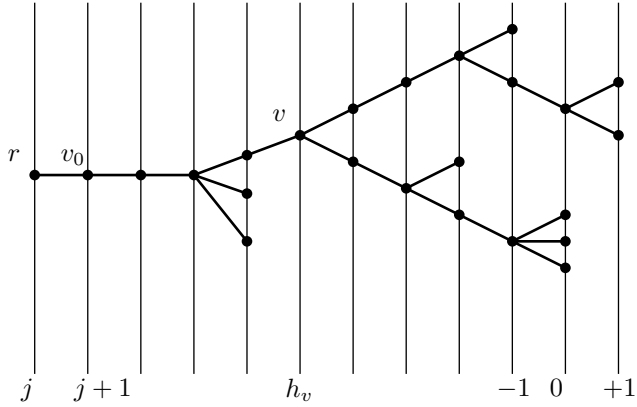
\centerline{Fig. 1: A tree $\t$ and its labels.}
\* 
\vglue.5truecm
We need some definitions and notations.
 
\0 1) Let us consider the family of all trees which can be constructed
by joining a point $r$, the {\it root}, with an ordered set of $n\ge 1$
points, the {\it endpoints} of the {\it unlabeled tree} (see Fig.
\graf(1)), so that $r$ is not a branching point. $n$ will be called the
{\it order} of the unlabeled tree and the branching points will be called
the {\it non trivial vertices}.
The unlabeled trees are partially ordered from the root to the endpoints in
the natural way; we shall use the symbol $<$ to denote the partial order.
 
Two unlabeled trees are identified if they can be superposed by a suitable
continuous deformation, so that the endpoints with the same index coincide.
It is then easy to see that the number of unlabeled trees with $n$ end-points
is bounded by $4^n$.
 
We shall consider also the {\it labeled trees} (to be called simply trees in
the following); they are defined by associating some labels with the unlabeled
trees, as explained in the following items.
 
\0 2) We associate a label $h\le 0$ with the root and we denote $\TT_{h,n}$ the
corresponding set of labeled trees with $n$ endpoints. Moreover, we introduce
a family of vertical lines, labeled by an an integer taking values in
$[h,2]$, and we represent any tree $\t\in\TT_{h,n}$ so that, if $v$ is an
endpoint or a non trivial vertex, it is contained in a vertical line with
index $h_v>h$, to be called the {\it scale} of $v$, while the root is on the
line with index $h$. There is the constraint that, if $v$ is an endpoint,
$h_v>h+1$.
 
The tree will intersect in general the vertical lines in set of
points different from the root, the endpoints and the non trivial vertices;
these points will be called {\it trivial vertices}. The set of the {\it
vertices} of $\t$ will be the union of the endpoints, the trivial vertices
and the non trivial vertices.
Note that, if $v_1$ and $v_2$ are two vertices and $v_1<v_2$, then
$h_{v_1}<h_{v_2}$.
Given a vertex $v$, which is not an endpoint, $\xx_v$ will denote the family
of all space-time points associated with one of the endpoints following $v$.
Moreover, there is only one vertex immediately following
the root, which will be denoted $v_0$ and can not be an endpoint;
its scale is $h+1$. 
Finally, if there is only one endpoint, its scale must be equal to $+2$ or
$h+2$.
 
\0 3) With each endpoint $v$ of scale $h_v=+2$ we associate one of the three
contributions to $\VV(\psi^{(\le 1)})$, written as in \equ(v) and a set
$\xx_v$ of space-time points, the corresponding integration variables.
With each endpoint $v$ of scale $h_v\le 1$ we associate one of 
local terms in $\LL\VV^{(h_v-1)}$ \equ(4.18); we will say that 
the endpoint is of type $g_1$, $g_2$ and so on depending
on the term we associate to it. 

Moreover, we impose the constraint that, if $v$ is an endpoint and $\xx_v$ is
a single space-time point (that is the corresponding term is local),
$h_v=h_{v'}+1$, if $v'$ is the non trivial vertex immediately preceding $v$.
 
\0 4) If $v$ is not an endpoint, the {\it cluster } $L_v$ with frequency $h_v$
is the set of endpoints following the vertex $v$; if $v$ is an endpoint, it is
itself a ({\it trivial}) cluster. The tree provides an organization of
endpoints into a hierarchy of clusters.
 
 
\0 5) We introduce a {\it field label} $f$ to distinguish the field variables
appearing in the terms associated with the endpoints as in item 3);
the set of field labels associated with the endpoint $v$ will be called $I_v$.
Analogously, if $v$ is not an endpoint, we shall
call $I_v$ the set of field labels associated with the endpoints following
the vertex $v$; $\xx(f)$, $\s(f)$ and $\o(f)$ will denote the space-time
point, the $\s$ index and the $\o$ index, respectively, of the
field variable with label $f$.
 
If $h\le 0$,
the effective potential can be written in the following way,see [BM1]:
$$\VV^{(h)}(\sqrt{Z_h}\psi^{(\le h)}) + L\b \tilde E_{h+1}=
\sum_{n=1}^\io\sum_{\t\in\TT_{h,n}}
V^{(h)}(\t,\sqrt{Z_h}\psi^{(\le h)})\Eq(3.27)\;,$$
where, if $v_0$ is the first vertex of $\t$ and $\t_1,..,\t_s$ ($s=s_{v_0}$)
are the subtrees of $\t$ with root $v_0$,\\
$V^{(h)}(\t,\sqrt{Z_h}\psi^{(\le h)})$ is defined inductively by the relation
$$\eqalign{
&\qquad V^{(h)}(\t,\sqrt{Z_h}\psi^{(\le h)})=\cr
&{(-1)^{s+1}\over s!} \EE^T_{h+1}[\bar
V^{(h+1)}(\t_1,\sqrt{Z_{h}}\psi^{(\le h+1)});..; \bar
V^{(h+1)}(\t_{s},\sqrt{Z_{h}}\psi^{(\le h+1)})]\;,\cr}\Eq(3.28)$$
and $\bar V^{(h+1)}(\t_i,\sqrt{Z_{h}}\psi^{(\le h+1)})$
 
\0 a) is equal to $\RR\hat \VV^{(h+1)}(\t_i,\sqrt{Z_{h}}\psi^{(\le h+1)})$ if
the subtree $\t_i$ is not trivial;
 
\0 b) if $\t_i$ is trivial and $h\le -1$, it
is equal to one of the terms in $\LL\VV^{(h+1)}$ 
\equ(4.18)or,
if $h=0$, to one of the terms contributing to $\hat\VV(\psi^{\le 1})$ \equ(v).
 
It is then easy to get, by iteration of the previous procedure, a simple
expression for $V^{(h)}(\t,\sqrt{Z_h}\psi^{(\le h)})$, for any
$\t\in\TT_{h,n}$.
 
We associate with any vertex $v$ of the
tree a subset $P_v$ of $I_v$, the {\it external fields} of $v$.
These subsets must satisfy various constraints. First of all, if $v$ is not
an endpoint and $v_1,\ldots,v_{s_v}$ are the vertices immediately following it,
then $P_v \subset \cup_i P_{v_i}$; if $v$ is an endpoint, $P_v=I_v$. We shall
denote $Q_{v_i}$ the intersection of $P_v$ and $P_{v_i}$; this definition
implies that $P_v=\cup_i Q_{v_i}$. The subsets $P_{v_i}\bs Q_{v_i}$,
whose union will be made, by definition, of the {\it internal fields} of $v$,
have to be non empty, if $s_v>1$.
 
Given $\t\in\TT_{h,n}$, there are many possible choices of the subsets $P_v$,
$v\in\t$, compatible with all the constraints; we shall denote $\PP_\t$ the
family of all these choices and $\bP$ the elements of $\PP_\t$. Then we can
write
$$V^{(h)}(\t,\sqrt{Z_h}\psi^{(\le h)})=\sum_{\bP\in\PP_\t}V^{(h)}(\t,\bP)
\;;\Eq(3.36)$$
Calling $W^{(h)}_{\t,\bP}$ the kernels of 
$V^{(h)}(\t,\bP)$ (see \equ(2.61a)) and
repeating the analysis in \S 3 of 
[BM1] one gets the following bound (analogous to (3.105) of [BM1])
$$\eqalign{
&\int d\xx_{v_0} |W^{(h)}_{\t,\bP}(\xx_{v_0})|\le
C^n L\b\e_h^n \g^{-h D_k(P_{v_0})}\;\cdot\cr
&\cdot\; \prod_{v\,\hbox{\ottorm not e.p.}} \chi(P_v)
\left\{ {1\over s_v!}
C^{\sum_{i=1}^{s_v}|P_{v_i}|-|P_v|}
\Big(Z_{h_v}/Z_{h_v-1}\Big)^{|P_v|/2}
\g^{-[-2+{|P_v|\over 2}+z(P_v)]}\right\}\;,\cr}\Eq(3.105)$$
where $z(P_v)=2$ if $|P_v|=2$ and $z(P_v)=1$ if $|P_v|=1$
and $||\sum_{f\in P_v}\e(f)\o(f)p_F||_{T^1}=0$; 
moreover $\chi(P_v)$
are defined so that
$\chi(P_v)=0$ if $|P_v|=4$ , $h_v\le \bar h$ and 
$||\sum_{f\in P_v}\e(f)\o(f)p_F||_{T^1}\not=0$,
and $\chi(P_v)=1$ otherwise. 

We call $2-{|P_v|\over 2}-z(P_v)$ the 
{\it dimension} of the vertex $v$ in the tree.
If no renormalization is defined $\RR=1$ then one gets a similar bound
with $z_v(P_v)=0$. Hence if $\RR=1$ the vertices $v$ with $|P_v|=4$
have vanishing dimension ({\it marginal terms}) while if 
$|P_v|=2$ they have positive dimension ({\it relevant terms}).
The presence of the $\chi$-functions in \equ(3.105)
is easily understood by noting
that one can insert freely such $\chi$ functi
ons in momentum space, 
then one passes to coordinate space and make bounds using 
the Gram-Hadamard inequality as in [BM1]. 

For any $v$ such that $h_v\le\bar h$ it holds  
$-2+{|P_v|\over 2}+z(P_v)\ge 1$, that is the dimension is negative, 
while if $h_v\ge\bar h$ it holds $-2+{|P_v|\over 2}+z(P_v)\ge 0$.

We have to perform the
sums over $\t$ and $\bP$. The number of unlabeled trees is $\le
4^n$; fixed an unlabeled tree, the number of terms in the sum over the
various labels of the tree is bounded by $C^n$, except the sums over the scale
labels and the sets $\bP$. 
 
In order to bound the sums over the scale labels and $\bP$ we first use
the inequality, for a constant $0<c<1$
$$\eqalign{
&\prod_{v\,\hbox{\ottorm not e.p.}} \chi(P_v)
\Big(Z_{h_v}/Z_{h_v-1}\Big)^{|P_v|/2}
\g^{-[-2+{|P_v|\over 2}+z(P_v)]}\le\cr
&\le [\prod_{\tilde v} (\chi(h_{\tilde v}\le \bar h)\g^{-c (h_{\tilde v}-h_{\tilde v'})}
+\chi(h_{\tilde v}\ge \bar h))]
[\prod_{v\,\hbox{\ottorm not e.p.}}\chi(|P_v|>4)\g^{-{|P_v|\over 40}}]\;,\cr}\Eq(3.111)$$
where $\tilde v$ are the non trivial vertices, and $\tilde v'$ is the
non trivial vertex immediately preceding $\tilde v$ or the root. Then
it holds that, noting the the number of nontrivial vertices is bounded by $n$
$$\sum_{ \{h_{\tilde v}\}}
[\prod_{\tilde v} 
(\chi(h_{\tilde v}\le \bar h)\g^{-c (h_{\tilde v}-h_{\tilde v'})}
+\chi(h_{\tilde v}\ge \bar h))]
\le C^n |\bar h|^{\a n}\Eq(3.11a)$$
for some numerical constant $\a$.
Finally the sum over $\bP$ can be done as described in [BM1].
\qed 
\vskip.5cm
{\bf Remark} 

By \equ(3.111) we get also
that the bound for a tree $\t\in \TT_{h,n}$ with at least a vertex at scale $k$
improves by a factor $\g^{\th(h-k)}$; this property is called {\it short memory property}.
\vskip.5cm
\section(4, The flow equation)
\vskip.5cm
\* \sub(4.1) {\it Second order analysis.}\*

By the iterative integration procedure seen in the previous section
it follows that the running coupling constants verify a recursive relation whose
r.h.s. is called {\it Beta function}:
$${Z_{h-1}\over Z_h} = 1+ z_h(v_h,..,v_1)\Eq(b1)$$
$$\n_{h-1}=\g\n_h+\b_\n^{(h)}(v_h,..,v_1)\Eq(b2)$$ 
$$\d_{h-1}=\d_h+\b_\d^{(h)}(v_h,..,v_1)\Eq(b3)$$ 
$$g_{i,h-1}=g_{i,h}+\b_{g,i}^{(h)}(v_h,..,v_1)\Eq(b4)$$
with $i=(1,2,3)$. The above equations are also called flow equations.
The functions $z_h,\b_\n^{(h)},\b_\d^{(h)}
\b_{g,i}^{(h)})$ are expressed by the tree
expansion seen in \S 3 (for details, see [BM1]). 
The contribution to $\b_{g,1}^{(h)}$ from the trees
with two end-points associated to the quartic running coupling constants is given by,
if $\int d\rr=\int_{-\b/2}^{\b/2} dr_0 \sum_{r\in\L}$
$$\sum_{k\le h} 4 \int d\rr g^{(k)}_\o(\rr) g^{(h)}_{-\o}(-\rr) g_{1,k} g_{1,k}
= 4\int d\rr \sum_{k=h,h+1} g^{(k)}_\o(\rr) g^{(h)}_{-\o}(-\rr)  g_{1,h}g_{1,k}\Eq(sec)$$
Using that
$g_{1,h}-g_{1,h+1}=O(v^2_h)$ and \equ(dec), and computing the equation analogue
to \equ(sec) for $g_{2,h-1}$ and $g_{4,h-1}$ we get
that $g_{i,h}$ verify the following equations
$$g_{1,h-1}=g_{1,h}-a g_{1,h}^2+O(\bar v_h^2\g^{\th h})+O(\bar v_h^3)$$
$$g_{2,h-1}=g_{2,h}-{a\over 2} g_{1,h}^2+O(\bar v_h^2\g^{\th h})+O(\bar v_h^3)\Eq(trunc)$$
$$g_{4,h-1}=g_{4,h}+O(\bar v_h^2\g^{\th h})+O(\bar v_h^3)$$
with $a$ a positive constant, given by 
$$a=a_1+a_2=4 \int d\rr [g^{(h)}_{L,\o}(\rr) g^{(h)}_{L,-\o}(\rr)
+g^{(h)}_{L,\o}(\rr) g^{(h+1)}_{L,-\o}(\rr)]
\Eq(att)$$
If we neglect the cubic contributions $O(\bar v_h^3)$ it is easy to see that 
the flow is bounded (in sense
that the quartic running coupling remain smaller than $O(U)$ for any
$h$) if $U>0$; in the general case in which the interaction is non local
the conditions is
$g_{1,0}=U v(2 p_F)+O(U^2)>0$. Of taking into account all higher order terms could 
destroy such behavior; aim of the following sections is to prove that
also taking into account the full Beta function
the quartic running coupling remain smaller than $O(U)$. 
\* 
\sub(4.1) {\it Beta function decomposition.}\*

We have two free parameters pot our disposal, $\n$ and $\d$; we will show that we can
fix them so that $\n_h=O(U^2 \g^{\t h})$ and  $\d_h=O(U^2 \g^{\t h})$. 
We fix then our attention on the flow equation for $g_{1,h},g_{2,h} , g_{4,h}$.

More explicitly \equ(b4) can be written as
$$g_{1,h-1}=g_{1,h}-g_{1,h}[a_1 g_{1,h}+a_2 g_{1,h+1}]+
G_{h}^1(g_h,..,g_0)+\sum_{k,k'} g_{1,k}g_{1,k'} H_{h,k,k'}^1(v_h,..,v_0)+R_h^1(v_h,...v_1)$$
$$g_{2,h-1}=g_{2,h}-{1\over 2} g_{1,h}[a_1 g_{1,h}+a_2 g_{1,h+1}]+\b^{2}_h(\m_h,..,\m_0)+
G_{h}^2(g_h,..,g_0)+\sum_{k,k'} g_{1,k}g_{1,k'} H_{h,k,k'}^2(v_h,..,v_0)+R^2_h(v_h,...v_1)\Eq(f1)$$
$$g_{4,h-1}=g_{2,h}+\b^{4}_h(\m_h,..,\m_0)+
G_{h}^4(g_h,..,g_0)+\sum_{k,k'} g_{1,k}g_{1,k'} H_{h,k,k'}^4(v_h,..,v_0)+R_h^4(v_h,...v_1)$$
where the following definitions are used:
\*
1)We write in \equ(b1) $z_k=z_k^1+z_k^2$, where 
$z_k^1$ is defined iteratively as the sum of all 
trees with only end-points at scale $\le 0$ and with propagators 
$g_L^{(k)}$, see \equ(dec), and in which 
${Z_{k'-1}\over Z_{k'}}$, $k'\ge k$ is replaced by $1+z_{k'}^1$.
\*
2) The functions $\b^{2}_h, 
\b^{4}_h, G_{h}^2,G_{h}^4, G_{h}^1, g_1 g_1 H^i$,with $i=1,2,4$ 
are the sum of all the trees with only end-points at scale $\le 0$ and with propagators 
$g_L^{(k)}$, see \equ(dec), and in which the factors
${Z_{k-1}\over Z_{k}}$, $k\ge h$ are replaced by $1+z_{k}^1$.
\*
3)The terms contributing to $\b^{2}_h, 
\b^{4}_h$ are by definition independent from
$g_{1,k}$,$k\ge h$.
\*
4)The terms contributing to $G^1_h,G^2_h,G^4_h$
by definitions depend linearly from 
$g_{1,k}$, that is they are vanishing if
$g_{1,k}=0$ for any $k$ and their second derivatives respect to
$g_{1,k}$ are also vanishing, while the first derivative are not vanishing.
\*
5)The terms at least quadratic in $g_1$ are included in
$\sum_{k,k'} g_{1,k}g_{1,k'} H_{h,k,k'}^i$
and by \equ(3.111) it holds
$$|H_{h,k,k'}^i|\le C \bar v_h \g^{\th(h-k)}\g^{\th(h-k')}\Eq(sm)$$
\*  
6)In
$R^{(i)}_h$ we include; terms depending from $\n_h$ or 
$\d_h$; terms with at least a propagator
$r_{1}^h(\xx-\yy)$, see \equ(dec); or terms with at least an 
endpoint at scale $1$.
\*
Note that the above decomposition is obtained by an analogous 
decomposition over trees,
so that the determinant bounds of \S 3 are still valid. 

In writing \equ(f1)
we have used that the beta function contributing to 
$g_1$ has at least a $g_1$;
in fact consider a contribution to the antiparallel part of $g_1$; 
it is not invariant under the transformation $\psi_{1,\s}^\pm\to e^{\pm \s}\psi_{1,\s}^\pm$
and $\psi_{-1,\s}^\pm\to \psi_{-1,\s}^\pm$ while the terms 
corresponding to $g_2$ and $g_4$ are invariant.

The flow given by \equ(f1) is very difficult to study; luckily dramatic 
cancellations appear, given by,
if $\bar g_h=\max_{k\ge h}
(|g^1_k|+|g^2_k|+|g^4_k|)$ and  $\bar\m_h=\max_{k\ge h} (|g^2_k|+|g^2_k|)$, the following result.
\*
{\cs Theorem 4 (Partial vanishing of the Beta function)} {\it The functions $\b^{2}_h, 
\b^{4}_h, G_{h}^2,G_h^4,G_{h}^1$, for $|v_h|\le\e$ are such that,
for a suitable constant $C$
$$|\b^{2}_h(\m_h,..,\m_h)|\le C\bar \m_h^2\g^{\th h}\quad |\b^{4}_h(\m_h,..,\m_h)|\le C\bar \m_h^2\g^{\th h}\Eq(1a)$$
$$|G_{h}^2(g_h,..,g_h)|\le  C\bar g_h^2\g^{\th h}
\quad |G_{h}^4(g_h,..,g_h)|\le  C\bar g_h^2\g^{\th h}\Eq(1b)$$
$$|G_{h}^1(g_h,..,g_h)|\le  C\bar g_h^2\g^{\th h}\Eq(1c)$$}
\*
The above lemma says that a dramatic cancellation happens in the series for
the above functions; each order is sum of many terms 
$O(1)$, but at the end the final sum is $O(\g^{\th h})$, that is
asymptotically vanishing. We call such property {\it partial
vanishing of the Beta function} (partial becouse the $O(g_1^2)$ terms are not vanishing).

Assuming the above lemma, which will proved in the following two sections
as consequence of suitable Ward identities, 
we can prove that the flow is bounded for any $g_{1,0}>0$.

We proceed in the following way.
We first {\it assign} a sequence 
$\un\defin\{\n_h\}_{h\le 1}$,
${\underline\d}\defin\{\d_h\}_{h\le 1}$ 
not necessarily solving the 
flow equation for $\n,\d$, but such that $|\n_h|, |\d_h|\le c U\g^{\th h}$, 
for any $h\le 1$. We then solve the flow equation for $g_{i,h}$,
parametrically in $\n,\d$, and show that, {\it for any sequence} $\un,\underline{\d}$
with the supposed property, the solution ${\underline g}(\un,{\underline\d})=\{g_{1,h}(\un,{\underline\d}),
g_{2,h}(\un,{\underline\d}),g_{4,h}(\un,{\underline\d})
\}_{h\le 1}$ 
exists and has good decaying properties. We finally fix the sequence $\un$ via
a convergent iterative procedure.

\vskip.5cm
{\cs Lemma 1.}{\it 
Assume that 
$|\n_h|, |\d_h|\le c U\g^{\th h}$ for any $h$.
For $U>0$ and small enough the flow is given by, for any $h$}
$$|g_{2,h}-g_{2,0}-g_{1,0}/2|\le U^{3/2}\quad |g_{4,h}-g_{4,0}|\le U^{3/2}
\quad 0<g_{1,h}\le {g_{1,0} \over 1-a/3 g_{1,0} h}\Eq(opp)$$
\vskip.5cm
{\it Proof.} By using that $|\n_h|, |\d_h|\le c U\g^{\th h}$
it holds that
$$|R^i_h|\le C U^2 \g^{\th h}\Eq(sm1)$$
It is convenient to introduce $\tilde g_{2,h}=2 g_{2,h}-g_{1,h}$;
then using \equ(1a) and \equ(sm1)
$$\tilde g_{2,h-1}=\tilde g_{2,h}+\sum_{k\ge h} D_{h,k}
+\sum_{k\ge h} (2 D^2_{h,k}-D^1_{h,k})+\sum_{k,k'}g_{1,k}g_{1,k'}\bar H_{h,k,k'}+\bar R_h \Eq(f22)$$
with 
$$D_{h,k}=\b^{2}_h(\m_h,...\m_h,\m_k,\m_{k+1},..,\m_0)-
\b^{2}_h(\m_h,...\m_h,\m_h,\m_{k+1},..,\m_0)\Eq(f22ww)$$
$$D^i_{h,k}=G^{i}_{h}(g_h,...g_h,g_k',g_{k'+1},..,g_0)-
G^{i}_{h}(g_h,...g_h,g_h,g_{k'+1},..,g_0)\quad i=1,2$$
and a similar equation for $g_{4,h}$; $\bar H_{h,k,k'}$ verifies \equ(sm),
$\bar R_h$ \equ(sm1) and 
$$|D_{h,k}|\le C \g^{-2\th(k-h)}U|g_h-g_k|\quad\quad
|D^i_{h,k}|\le C U \g^{-2\th(k-h)}|g_h-g_k|\Eq(ind2)$$
Assume that for $k> h$ 
%
$$0\le g_{1,k-1}\le {g_{1,0} \over 1-a/3 g_{1,0} (k-1)}\quad |g_{k-1}-g_{k}|\le [U^{5\over 4} \g^{\th k}+
[{g_{1,0} \over 1-a/3 g_{1,0} k}]^2]\Eq(ind1)$$ 

We have then to prove that such inequalities hold for $k=h-1$. 
Noting that
$$\sum_{k=h}^{-1}\g^{\th(h-k)}{1\over -k}={1\over -h}\sum_{k=h}^{-1}\g^{\th(h-k)}
+\sum_{k=h}^{-1}\g^{\th(h-k)}{(k-h)\over k h}\le {C_1\over -h}\Eq(oppo)$$
we obtain
$$\sum_{k,k'} g_{1,k}g_{1,k'} \bar H_{h,k,k'}^2\le C U g_{1,h}^2\Eq(oppj)$$
Moreover 
$$\sum_{k\ge h} |D_{h,k}|\le \sum_{k=-1}^h C C_1 U \g^{-2\th(k-h)}\sum_{k'=h}^k
|U^{5\over 4} \g^{\th k'}+
[{g_{1,0} \over 1-a/3 g_{1,0} k'}]^2|\le $$
$$C_2 C U (U^{5\over 4} \g^{\th h}+\sum_{k=-1}^h|k-h|\g^{2\th(h-k)}
[{g_{1,0} \over 1-a/3 g_{1,0} k}]^2)\Eq(lau)$$
and the last addend can be bounded by
$$\sum_{k=-1}^h\g^{\th(h-k)}{1\over k^2}\le 
 C_2  [{1 \over 1-a/3 g_{1,0} h}]^2\;.\Eq(bbl)$$
%
%
%
%
%
Then
by \equ(f22) we get
$$|\tilde g_{2,h-1}-\tilde g_{2,h}|\le C_3 (U^{2}\g^{\th h}+
U ({g_{1,0}\over 1-{a\over 3}g_{1,0}h})^2)\Eq(f22)$$
and
$$|\tilde g_{2,h-1}-\tilde g_{2,0}|\le C_3 (U^{2}\sum_{k=h}^0\g^{\th k}+\sum_k 
U [{g_{1,0} \over 1-a/3 g_{1,0} k}]^2\le U^{3/2}\Eq(ghh)$$
In the same way in the flow for $g_4$ we use that there are no second order contributions
quadratic in $g_{1,h}$.
Finally we write, using \equ(f1) and the short memory property
(namely that $\g^{\th(h-k)}g_{1,k}\le C g_{1,h}$)
$$g_{1,h-1}-g_{1,h}\le  -{a\over 3} g_{1,h} g_{1,h-1}
\Eq(oppo2)$$
or
$$g_{1,h-1}\le {g_{1,h}\over 1+{a\over 3}g_{1,h}}\Eq(oppo2)$$
and as ${x\over 1+x}$ is an increasing function and by induction 
$0<g_{1,h}\le {g_{1,0}\over 1-{a\over 3}g_{1,0}h}$
so that
$$g_{1,h-1}\le 
{g_{1,0} (1-{a\over 3}h g_{1,0})^{-1}
\over 1+{a\over 3} 
g_{1,0}(1-{a\over 3} h g_{1,0})^{-1}}\le 
{g_{1,0}\over 1-{a\over 3}g_{1,0}(h-1)}
\;.\Eq(oppo3)$$
Moreover $g_{1,h-1}=g_{1,h}(1+O(U))$
by \equ(f1), and $g_{1,h}>0$ so that $g_{1,h-1}>0$.
\qed
\vskip.5cm
\* \sub(4.4) {\it The choice of the counterterms.}\*

In the previous section we solve the flow equation 
for $g_{i,h}$ parametrically in any sequence $\un=\{\n_h\}_{h\le 1}$,
${\underline\d}=\{\d_h\}_{h\le 1}$ such that 
$|\n_h|\le c U\g^{\th h}$, $|\d_h|\le c U\g^{\th h}$
for any $h$. We show now that indeed we can choose $\n,\d$
so that $\un=\{\n_h\}_{h\le 1}$,
${\underline\d}=\{\d_h\}_{h\le 1}$ verify such a property.
\vskip.5cm
{\cs Lemma 2 }{\it
There exist sequences 
$\un=\{\n_h\}_{h\le 1}$,
${\underline\d}=\{\d_h\}_{h\le 1}$ 
such that $|\n_h|\le c U\g^{\th h}$, $|\d_h|\le c U\g^{\th h}$.}\\
\*
{\it Proof} It holds that
$$\b_{\d}^{(h)}=\b_{\d,a}^{(h)}+\b_{\d,b}^{(h)}\Eq(aa3)$$
where $\b_{\d,a}^{(h)}$ us given by a sum of trees with no end-points
$\n_k,\d_k$ and only propagators $g_{L,\o}^k$ \equ(dec); 
by the symmetry in the exchange
$x,x_0$ of $g_L$, and remembering that $\b_\d^{(h)}=\sum_\t[z(\t)-a(\t)]$
it holds that 
$$|\b_{\d,a}^{(h)}|\le C U \g^{2\th h}\Eq(aa2)$$ 
A similar decomposition can be done also for 
$$\b_{\n}^{(h)}=\b_{\n,a}^{(h)}+\b_{\n,b}^{(h)}\Eq(aa4)$$
again with
$$|\b_{\n,a}^{(h)}|\le C U \g^{2\th h}\Eq(aaa1)$$ 
by the parity property $g_{L,\o}^h(\xx,\yy)=-g_{L,\o}^h(\yy,\xx)$.
If we want to fix $\n,\d$ in such a way that $\n_{-\io}=\d_{-\io}=0$, we must have,
if $(\n_1,\d_1)=(\n,\d)$:
$$\n=-\sum_{k=-\io}^1\g^{k-2}\b^k_\n(g_k,\d_k,\n_k;\ldots;g_1,\d_1,\n_1)\;.\Eq(5.8a)$$
$$\d=-\sum_{k=-\io}^1\b^k_\d(g_k,\d_k,\n_k;\ldots;g_1,\d_1,\n_1)\;.
\Eq(5.8bb)$$
Note that in \equ(5.8a),\equ(5.8bb) $g_k\equiv g_k(\un, {\underline\d})$.

If we manage to fix $\n,\d$ as in \equ(5.8a), \equ(5.8bb) we also get:
$$\n_h=-\sum_{k\le h}\g^{k-h-1}\b^k_\n(g_k,\d_k,\n_k;\ldots;g_1,\d_1,\n_1)\;.
\Eq(5.8b)$$
$$\d_h=-\sum_{k\le h}\b^k_\d(g_k,\d_k,\n_k;\ldots;g_1,\d_1,\n_1)\;.
\Eq(5.8b)$$

Let $\MMM_\th$ be the space of sequences $\un=\{\n_{-\io},\ldots,\n_1\}$,
${\underline\d}=
\{\d_{-\io},..,\d_1\}$
with small $||\cdot||_\th$ norm, namely the space of sequences $\un$ 
satisfying:

$$|\d_k|\le \g^{\th k},|\n_k|\le \g^{\th k} $$
We look for a fixed point 
of the operator $\bT:\MMM_{\th}\to\MMM_{\th}$ defined as:
$$T(\n_h)=-\sum_{k\le h}\g^{k-h-1}\b^k_\n(g_k({\underline\d},\un),\n_k;\ldots;g_1,\n_1)
\,.\Eq(5.8c)$$
$$T(\d_h)=-\sum_{k\le h}\b^k_\d(g_k({\underline\d},\un),
\d_k,\n_k;\ldots;g_1,\d_1,\n_1)
\,.\Eq(5.8cs)$$

First note that, if $U$ is sufficiently small, then $\bT$ leaves 
$\MMM_\th$ invariant: in fact 
$$|(\bT \n)_h|\le \sum_{k\le h} 2 c_1 U\g^{\th k}\g^{k-h}\le 
c U \g^{\th h}\quad
|(\bT \d)_h|\le \sum_{k\le h} 2 c_1 U\g^{\th k}\le 
c U \g^{\th h}\Eq(aaa12)$$

Furthermore we find that 
$\bT$ is a contraction on $\MMM_\th$: in fact
$$\eqalign{&|(\bT \d)_h-(\bT\d')_h|\le \sum_{k\le h}
|\b^k_\d(g_k(\un,\d),\n_k,\d_k;
\ldots)-\b^k_\n(g_k(\un',\underline{\d}'),\n_k',\underline{\d}'_k;
\ldots)|\le\cr  
&\le c''U\g^{\th h}
[||\un-\un'||_\th+||\underline{\d}-\underline{\d}'||_\th]  \;.\cr}\Eq(5.8e)$$
and a similar equation holds for $\n$.
Then, a unique fixed point $\un^*, {\underline\d}$ for $\bT$ exists on $\MMM_\th$. 
\qed
\*
By the above Lemma we have found $\d(\tilde t, p_F, U),\n(\tilde t, p_F, U)$;
inserting them in \equ(bb) and using the implicit function theorem we get
$p_F(U,\m), \tilde t(U,\m)$.

Finally from an explicit second order computation we obtain that
$$z_h= a[g_{1,h}^2+g_{2,h}^2+g_{4,h}^2]+\b^{\ge 3}\Eq(ll5)$$
with $a>0$ is a suitable constant,
and using the previous results on the flow of $g_{i,h},\n_h,\d_h$
we get $\lim_{h\to\io} {Z_h\over \g^{\h h}}=1$, where $\h= a [g_{2,-\io}^2+g_{4,-\io}^2]+O(U^3)$.
\*

\vskip.5cm
\section(5, The reference model and proof of Theorem 4)
\vskip.5cm
\* \sub(5.1) {\it The model.}\*

In order to prove the partial vanishing of the Hubbard model Beta function
expressed by  
\equ(1a),

\equ(1b),\equ(1c) we introduce a {\it reference model}
written directly in terms of grassmann variables, with an ultraviolet cutoff
and an infrared cutoff $\g^h$ {\it with linear dispersion relation
and in the continuum}. We study the reference model by Renormalization Group
and we show that the Beta function of this model is asymptotically vanishing 
as a consequence of {\it Ward identities} due to the formal local chiral
gauge invariance (which is however broken
by the presence of cutoffs); then we prove that the Beta function of the reference
model coincides partly with the Beta function of the Hubbard model,so that
we can deduce the {\it partial vanishing} of the Hubbard model Beta function
from the vanishing of the reference model Beta function.

The partition 
function of the reference model is  
$$\int P_L(d\psi) e^{\VV_L}\Eq(mmnb)$$
where the propagator is 
$$g^h_{\o,L}(\xx-\yy)=
{1\over\b L}\sum_{\kk\in \DD_{L\b}} C^{-1}_{h,0}(\kk) {e^{-i\kk(\xx-\yy)}\over -ik_0+\o k}\Eq(ppll)$$
with $C^{-1}_{h,0}(\kk)=\sum_{k=-\io}^0 f^k(k_0,k)$ and
$$\VV_L=\sum_\o \int_{-\b/2}^{\b/2} dx_0  
\int_{-L/2}^{L/2} dx
[g_2^o \psi^{+}_{\xx,\o,\s}\psi^-_{\xx,\o,\s}
\psi^+_{\xx,-\o,\s}\psi^-_{\xx,-\o,\s}+$$
$$g_2^p \psi^+_{\xx,\o,\s}\psi^-_{\xx,\o,\s}
\psi^+_{\xx,-\o,-\s}\psi^-_{\xx,-\o,-\s}+g_4
\psi^+_{\xx,\o,\s}\psi^-_{\xx,\o,\s}
\psi^+_{\xx,\o,-\s}\psi^-_{\xx,\o,-\s}]\Eq(ll)$$
Note that the model is not 
$SU(2)$ invariant, as the interaction
depends from the spin if $g_2^o\not=g_2^p$. 

The Grassmann integration can be done by a multiscale analysis 
essentially identical to the one described in \S 3; however the symmetries
of the interaction imply that 
the local part of the effective
potential \equ(4.18) is replaced by 
$$\LL\VV^j_L=\sum_\o \int_{-\b/2}^{\b/2} dx_0  
\int_{-L/2}^{L/2} dx
[\tilde g_{2,j}^p \psi^+_{\xx,\o,\s}\psi^-_{\xx,\o,\s}
\psi^+_{\xx,-\o,\s}\psi^-_{\xx,-\o,\s}+\tilde g_{2,j}^o
\psi^+_{\xx,\o,\s}\psi^-_{\xx,\o,\s}
\psi^+_{\xx,-\o,-\s}\psi^-_{\xx,-\o,-\s}]\Eq(5.1)$$
$$+\tilde g_{4,j} [
\psi^+_{\xx,\o,\s}\psi^-_{\xx_2,\o,\s}
\psi^+_{\xx,\o,-\s}\psi^-_{\xx,\o,-\s}]$$
Note in fact that the analogue of $\n_h,\d_h$ are vanishing by (in the limit $L,\b\to\io$)
parity and invariance in the exchange $(x,x_0)\to (x_0,x)$
; moreover the reference model is invariant under the total gauge transformation
$\psi_{\xx,\o,\s}^\pm \to e^{\pm\a_{\o,\s}}\psi_{\xx,\o,\s}^\pm$ for any values of 
$\a_{\o,\s}$, so that terms of the form 
$\psi^+_{\o,\s}\psi^-_{,-\o,\s}\psi^+_{-\o,-\s}\psi^-_{\o,-\s}$
cannot be generated in the integration procedure
as they violate such symmetry. Note also that,
due to the compact support of the cutoff in \equ(ppll), 
the running coupling constants at scale $k>h$ of the theory with infrared
cutoff $\g^h$ or $0$ are identical.

It is easy to verify that a tree expansion similar to the one described in \S 3.5
holds also for the reference model, and that the analogue of Theorem 3 holds also in this case. 
We will prove in \S 6 the following Lemma.
\*
{\cs Lemma 3}{\it Assume that $\bar g=max(|g_{2}^o|, 
|g_{2}^p|,|g_{4}|)$ is small enough; then 
for any integer $j\le 0$, in the limit $h\to-\io$ for a suitable constant $C$
$$|\tilde g_{2,j}^o-g_{2}^o|\le C\bar g^2\quad 
|\tilde g_{2,j}^p-g_{2}^p|\le C\bar g^2
\quad |\tilde g_{4,j}-g^{4}_p|\le C\bar g^2\Eq(5)$$
Moreover $\tilde g_{2,j}^o, \tilde g_{2,j}^p, \tilde g_{4,j}$
have a limit as $h\to-\io$.
}
\*
It is an immediate corollary of Lemma 3 and Theorem 3 that 
$\vec v^L_k=(\tilde g_{2,k}^o, \tilde g_{2,k}^p,\tilde g_{4,k}) $ are analytic functions
of $\vec v_1=(g_2^p,g_2^p,g_4)$ around $(0,0,0)$.
Note that analyticity in the coupling 
around the origin holds for the reference model and {\it not} for the 
Hubbard model.
Finally the analogue of the flow equations \equ(f1) is 
given by, if $\vec v^L_k=(\tilde g_{2,k}^o, \tilde g_{2,k}^p,
\tilde g_{4,k})$
$$\tilde g_{2,j-1}^o=\tilde g_{2,j}^o+\tilde\b^{2,o}_j(\vec v^L_j,..,\vec v^L_0)$$
$$\tilde g_{2,j-1}^p=\tilde g_{2,j}^p+\tilde \b^{2,p}_j(\vec v^L_j,..,\vec v^L_0)\Eq(5.2a)$$
$$\tilde g_{4,j-1}=\tilde g_{4,j}+\tilde \b^{4}_j(\vec v^L_j,..,\vec v^L_0)$$
We can rewrite the above equations as,for $j> h$
$$\tilde g_{2,j-1}^o=g_{2,j}^o+\tilde \b^{2,o}_j(\vec v^L_j,..,\vec v^L_j)+
\sum_{k>j} D^{2,o}_{j,k}$$
$$\tilde g_{2,j-1}^p=\tilde g_{2,j}^p+\tilde \b^{2,p}_j(\vec v^L_j,..,\vec v^L_j)+\sum_{k>j} 
\tilde D^{2,o}_{j,k}\Eq(5.44d)$$
$$\tilde g_{4,j-1}=\tilde g_{4,j}+\tilde \b^{4}_j(\vec v^L_j,..,\vec v^L_j)+
\sum_{k>j} D^{2,o}_{j,k}$$
with,for $\a=(2,o),(2,p),4$
$$\tilde D_{j,k}^{\a}=\tilde\b^{\a}_j(\vec v^L_j,...,\vec v^L_j,\vec v^L_k,\vec 
v^L_{k+1},..,\vec v^L_0)-
\tilde\b^{\a}_j(\vec v^L_j,...,\vec v^L_j,\vec v^L_j,\vec v^L_{j+1},..,\vec v^L_0)\Eq(f22ww)$$

\vskip.5cm
\* \sub(5.2) {\it Vanishing of the reference model beta function.}\*

The Beta function is an analytic function of 
$\vec v^L_j$ and it can be written as, if $\a=(o,2),(p,2),4$ 
$$\tilde\b^{\a}_j(\vec v^L_j,...,\vec v^L_j)=\sum_{n_1,n_2,n_3} b^\a_{j,n_1,n_2,n_3}
(\tilde g_{2,j}^o)^{n_1} (\tilde g_{2,j}^p)^{n_2} (\tilde g_{4,j})^{n_3} $$
We define $n\equiv n_1+n_2+n_3$ and $\vec{n}=(n_1,n_2,n_3)$.
Note that
$$ b^\a_{j,n_1,n_2,n_3}= b^\a_{n_1,n_2,n_3}+O(\g^{\th j})$$
Consider  $b^\a_{j,n_1,n_2,n_3}$
and  $b^\a_{k,n_1,n_2,n_3}$ with $k<j$;
for any tree $\t$ contributing to $\b_k$ there is a tree
contributing to $\b_j$; in fact
we can perform a change of variables
in the propagator $g^i(\kk)$ 
respectively $\kk\to \g^{j}\bar\kk$ and $\kk\to \g^{k}\bar\kk$,
so that in one case the propagator is $f(\g^{-i+k}\bar\kk) D_\o^{-1}(\bar\kk)$
and in the other  $f(\g^{-i+j}\bar\kk) D_\o^{-1}(\bar\kk)=
f(\g^{-i+k}\g^{-k+j}\bar\kk) D_\o^{-1}(\bar\kk)$;
hence for each tree contributing
to $b_j$ there is a tree contributing to $b_k$, in which
the scale of each vertex is shifted by $j-k$; there are extra 
contributions to $b^k$ with at least a vertex with scale $>k-j$; such trees
have the root at scale $k$ so that, by the short memory property,
$b_j-b_k=O(\g^{\th j})$, with $0<\th<1$
a suitable constant, and taking the limit $k\to-\io$
we get $b_j=b+O(\g^{\th j})$.

We will prove the following result.
\*
\vskip.5cm
{\cs Lemma 4} {\it Assume that Lemma 3 holds; then for any
$(n_1,n_2,n_3)$
$$b^\a_{n_1,n_2,n_3}=0\Eq(1f)$$}
\*
{\it Proof.}
The proof is by
contradiction; 
assume that, for some $\vec{\bar n}=(\bar n_1,\bar n_2,\bar n_3)$ 
with $\bar n=\bar n_1+\bar n_2+\bar n_3$ 
$$\vec \b_{j}(\vec v^L_h,...,\vec v^L_h)=\vec b_{j,\vec n}
(\tilde g_{2,j}^o)^{\bar n_1} (\tilde g_{2,j}^p)^{\bar n_2} (\tilde g_{4,j})^{\bar n_3}  
+O(\vec v^L_j)^{\bar n+1},\Eq(5.60z)$$
with $\vec b_{\vec{\bar n}}$ a non vanishing vector, and that for all $n_1+n_2+n_3=n\le \bar
n-1$, $\vec b_{\vec{n}}$ is vanishing. 
From Theorem 3 and Lemma 3 
$\vec v^L_j$ are analytic functions
of $\vec v_1=(g_2^o,g_2^p,g_4)$, that is
$$\vec v^L_j=\vec v_1+\sum_{n\le \bar n} \vec c^{(j)}_{\vec n} 
(g_{2}^o)^{n_1} (g_{2}^p)^{n_2} (g_{4})^{n_3} +O((\vec v^L_1)^{\bar n+1})\Eq(5.78c)$$
and for any fixed $j$ the sequence $\vec c^j_{\vec n}$ is a bounded sequence.
Inserting \equ(5.78c) in the Beta function, using analyticity and 
equating the coefficient of 
$(g_{2}^o)^{n_1} (g_{2}^p)^{n_2} (g_{4})^{n_3}$
with $n_1+n_2+n_3=n \le \bar n-1$ we get
$$\vec c^{(j-1)}_{\vec n} = \vec c^{(j)}_{\vec n}+ 
\sum_{k=j+1}^0  \vec d_{j,k}^{\vec n}+O(\g^{\th j})\Eq(5.78d)$$
where the last sum represents the contribution of
$\vec D_{j,k}$, so that
$$|\vec d_{j,k}^{\vec n}|\le \g^{-\th(k-j)} D_{\bar n} \sup_{2\le m\le n-1}
|\vec c_{\vec m}^{(j)}-\vec c_{\vec m}^{(k)}|\Eq(5.79a)$$
where we have used that $\vec D_{j,k}$ is at least quadratic in the running coupling 
constants, and $D_{\bar n}$ is a suitable constant (in $j$). Note that
$$|\vec c^{(j-1)}_{\vec n} -\vec c_{\vec n}|\le \bar D_{\bar n} 
\sum_{j'=-\io}^j[
\sum_{k=j'+1}^0  \g^{-\th(k-j')}\sup_{2\le m\le n-1}
|\vec c_{\vec m}^{(j')}-\vec c_{\vec m}^{(k)}|+ \g^{\th j'})]\Eq(5.78k)$$
The above inequality implies by induction that,for $n\le \bar n-1$ 
$$\sup_{2\le m\le n}|\vec c^{(k)}_{\vec m} -\vec c_{\vec m}|\le 
C^n \g^{{\th\over 2} k}\Eq(ind)$$
for a suitable $C$;
assume in fact that it is true for $k\ge j$ and from \equ(5.78k) 
we get
$$|\vec c^{(j-1)}_{\vec n} -\vec c_{\vec n}|\le C^{n-1}\bar D_{\bar n} 
\sum_{j'=-\io}^j[
\sum_{k=j'+1}^0 \g^{-\th(k-j')}(\g^{{\th\over 2} k}+\g^{{\th\over 2} j'})
+\g^{\th j'}]\le K C^{n-1}\bar D_{\bar n}\g^{{\th\over 2} (j-1)}\Eq(5.78l)$$
so that \equ(ind) holds for $j-1$ if $C\ge K \bar D_{\bar n}$.
On the other hand \equ(ind)
implies 
$$|\vec d_{j,k}^{\vec n}|\le \bar C^n \g^{{\th\over 2}(j-1)}\Eq(ind 1)$$
Writing now the analogous of \equ(5.78d) for $n=\bar n$ we get
$$\vec c^{(j-1)}_{\vec{\bar n}} = \vec c^{(j)}_{\vec{\bar n}}+\vec b_{j,\vec n}+ 
\vec d_{j,k}^{\vec{\bar n}}\Eq(5.78d)$$
which can be rewritten as
$$\vec c^{(j-1)}_{\vec{\bar n}}=\vec c^{(j)}_{\vec{\bar n}}+
\vec b_{\vec{\bar n}}+
O(\g^{{\th\over 2} j})\Eq(jgvvs)$$
so that $\vec c^{(j)}_{\vec{\bar n}}$
is necessarily a
diverging as $j\to\io$, and this is a contradiction.
\qed
\vskip.5cm
\* \sub(5.3) {\it Partial vanishing of the Hubbard model
Beta function
(Proof of Theorem 4).}\*

We compare the Beta functions
of the reference model with the function appearing
in the flow equations of the Hubbard model.
\*
A)Let us start considering first the reference model in 
the spin symmetric case, that is if $g_{2,0}^o=g_{2,0}^p$.
In such a case by the same arguments used in Appendix , 
for any $k$ $g_{2,k}^o=g_{2,k}^p$
and $\b_{2,k}^o=\b_{2,k}^p$, so that the flow equation \equ(5.2a)
reduces to  
$$\tilde g_{2,h-1}=\tilde g_{2,h-1}+\tilde\b^{2}_h(\tilde g_{2,h},\tilde g_{4,h};..,\tilde g_{2,0},\tilde g_{4,0})$$
$$\tilde g_{4,h-1}=\tilde g_{4,h-1}+\tilde \b^{4}_h(\tilde g_{2,h},\tilde g_{4,h};..;\tilde g_{2,0},\tilde g_{4,0})$$
It holds that the functions $\tilde\b^{2}_h$ and $\tilde \b^{4}_h$
essentially coincide with the functions $\b^2_h,\b^4_h$ of the Hubbard model
defined in \equ(f1); that is, if $\m_h=(g_{2,h},g_{4,h})$,for a suitable constant $C$
$$|\tilde\b^{2}_h(\m_h,..,\m_h)-\b^{2}_h(\m_h,..\m_h)|
\le C\m_h^2 \g^{\th h}\Eq(mmq)$$
$$|\tilde\b^{4}_h(\m_h,..,\m_h)-\b^{4}_h(\m_h,..\m_h)|
\le C\m_h^2 \g^{\th h}\Eq(mmq1)$$
The above equations prove \equ(1a).
In order to prove \equ(mmq) and \equ(mmq1) we note that
by definition the only difference between $\tilde\b^{2}_h,
\tilde\b^{4}_h$ and $\b^{2}_h,
\b^{4}_h$ is that in one case the model is defined on the continuum
and in the other case on the lattice. 
In momentum representation this means that
the delta functions in $\tilde\b$ are defined as $L\b \d_{k,0}\d_{k_0,0}$
while in $\b$ are defined as in \equ(2.62).
The difference of the two delta functions
slightly affects the non local terms on any scale,
hence it affects the beta function; however, it is easy to show that this is a
negligible phenomenon. Let us consider in fact a particular tree $\t$ and a
vertex $v\in\t$ of scale $h_v$ with $2n$ external fields of space momenta
$k'_r$, $r=1,\ldots,2n$; the conservation of momentum implies that
$\sum_{r=1}^{2n}\s_r k'_r=2\p m$, with $m=0$ in the continuous model, but
$m$ arbitrary integer for the lattice model. On the other hand, $k'_r$ is of
order $\g^{h_v}$ for any $r$, hence $m$ can be different from $0$ only if $n$
is of order $\g^{-h_v}$.
Since the number of endpoints following a vertex with $2n$ external
fields is greater or equal to $n-1$ and there is a small factor (of order
$\m_h$) associated with each endpoint, we get an improvement, in the bound of
the terms with $|m|>0$, with respect to the others, of a factor
$\exp(-C\g^{-h_v})$.
Hence it is easy to show that the
difference between the two beta functions is of order $\m_h^2\g^{\th h}$.
\*
B)In order to prove \equ(1b) we consider the reference model with 
$g_{2,0}^0\not= g_{2,0}^p$,
so that there are three independent running coupling constants.
We have seen that, for $\a=(2,p),(2,o),4$ 
$$\tilde\b^{\a}_h(v^L_h,..v^L_h)=\sum_{n_1,n_2,n_3} 
b_{h,n_1,n_2,n_3}^{\a} [g_{2,h}^o]^{n_1}
[g_{2,h}^p]^{n_2}[g_{4,h}]^{n_3}\Eq(ss1)$$
On the other hand we can write the functions $G^\a_h$ \equ(f1)
in the Hubbard model, $\a=(2o),(2p),4$, as
$$G^\a_h=\sum_{m_2,m_3} c_{h,1,n_2,n_3}^{\a} [g_{1,h}] [g_{2,h}]^{m_2}
[g_{4,h}]^{m_3}\Eq(ss2)$$
The coefficients $c_{h,1,n_2,n_3}^{\a}$ are given by sum of trees
(or product of trees, for the presence of the $z_k^1$ terms) 
with (in total) one end-point $g_1$, $m_2$ 
end-points $g_2$ and $m_3$ 
end-points $g_4$; the SU(2) invariance of the Hubbard model implies that
$G^{2o}_h=G^{2p}_h$.
To $g_1$ and $g_2$ correspond two terms,
the parallel or antiparallel part, 
see \equ(bb1), 
and we can associate to the endpoints of
the trees contributing to 
$c_{h, 1,m_2,m_3}^{\a}$ 
an extra index distinguishing the parallel or antiparallel 
part; then we can write
$$c_{h,1,m_2,m_3}^{\a}=\sum_{m_1^o+m_1^p=1}
\sum_{m_2^o+m_2^p=m_2}
c_{h,m_1^o,m_1^p,m_2^o,m_2^p,m_3}^{\a}\Eq(ss4)$$
%
%
It holds that
$$c_{h,1,m_2,m_3}^{\a}=
\sum_{m_2^o+m_2^p=m_2}
c_{h,0,1,m_2^o,m_2^p,m_3}^{\a}\Eq(ss7)$$
that is only the spin parallel part 
of $g_1$ can contribute to $G^2_h$ or $G^4_h$;
in fact making the
the global gauge transformation 
$\psi^\pm_{1,\s}\to e^{i\s}\psi^\pm_{1,\s}$ 
and  $\psi^\pm_{-1,\s}\to \psi^\pm_{-1,\s}$, the antiparallel part is not invariant,
while the spin parallel (and the $g_2$, $g_4$ interactions) are invariant.

Finally note that the spin parallel $g_1$ interaction
is equal (up to a sign) to the spin parallel $g_2$ interaction, so that, for 
$\a=(2o),(2p),4$
$$c_{0,1,m_2^o,m_2^p,m_3}^{\a}=-b_{m_2^o,m_2^p+1,m_3}^{\a}=0\Eq(llm)$$
\*
C)It remains to consider \equ(1c); we can consider equivalently 
the contribution to the spin parallel 
or the spin antiparallel, as they are equal 
by $SU(2)$ invariance of the Hubbard model, that is
$ G^{1o}_h=G^{1p}_h$.
We consider the spin parallel part
and we can write
$$G^{1p}_h=\sum_{m_2,m_3} c_{h,1,m_2,m_3}^{1p} [g_{1,h}] [g_{2,h}]^{m_2}
[g_{4,h}]^{m_3}\Eq(dfss)$$
with 
$$c_{1,m_2,m_3}^{1p}=\sum_{m_1^o+m_1^p=1}
\sum_{m_2^o+m_2^p=m_2}
c_{m_1^o,m_1^p,m_2^o,m_2^p,m_3}^{1p}$$
The single $g_1$ interaction cannot be antiparallel, 
again because making the
global gauge transformation 
$\psi^\pm_{1,\s}\to e^{i\s}\psi^\pm_{1,\s}$ 
and  $\psi^\pm_{-1,\s}\to \psi^\pm_{-1,\s}$, the antiparallel part is not invariant,
while the spin parallel (and the $g_2$, $g_4$ interactions) are invariant.
Hence
$$c_{1,m_2,m_3}^{1p}=
\sum_{m_2^o+m_2^p=m_2}
c_{0,1,m_2^o,m_2^p,m_3}^{1p}\Eq(dds)$$
and
$$c_{0,1,m_2^o,m_2^p,m_3}^{1p}=b^{2p}_{m_2^o,m_2^p+1,m_3}=0\Eq(jjk)$$
as the contribution $(1p)$ and $(2p)$ are identical.
\vskip.5cm
\section(6, Ward identities for the reference model: Proof of Lemma 3)
\vskip.5cm
\* \sub(5.2a){\it Dyson equations}

Let us now prove \equ(5), extending the analysis in [BM2], [BM3], [BM4] to the spinning case. 
We derive a number of Dyson equations relating some Schwinger functions of the reference
model. 
Let us start from, if $\r_{\pp,\o,\s}=
{1\over\b L}\sum_\kk \psi^+_{\kk,\o,\s}\psi^-_{\kk-\pp,\o,\s}$
$$<\psi^+_{\kk_1,+,\s}\psi^-_{\kk_2,+,\s}\psi^+_{\kk_3,-,-\s}\psi^-_{\kk_4,-,-\s}>_T=$$
$$g_{-}(\kk_4) \{G^2_{-}(\kk_3)[g_2^o <\psi^+_{\kk_1,+,\s}\psi^-_{\kk_2,+,\s} \r_{\kk_1-\kk_2,+,\s}>_T$$
$$+g_2^p <\psi^+_{\kk_1,+,\s}\psi^-_{\kk_2,+,\s} \r_{\kk_1-\kk_2,+,-\s}>_T$$
$$+g_4 <\psi^+_{\kk_1,+,\s}\psi^-_{\kk_2,+,\s} \r_{\kk_1-\kk_2,-,-\s}>_T]$$
$$+\int d\pp [g_2^o <\psi^+_{\kk_1,+,\s}\psi^-_{\kk_2,+,\s}
\psi^+_{\kk_3,-,\s}\psi^-_{\kk_4-\pp,-,-\s} \r_{\pp,+,\s}>_T+\Eq(b1)$$
$$\int d\pp g_2^p <\psi^+_{\kk_1,+,\s}\psi^-_{\kk_2,+,\s}
\psi^+_{\kk_3,-,\s}\psi^-_{\kk_4-\pp,-,-\s} \r_{\pp,+,-\s}>_T+$$
$$\int d\pp g_4 <\psi^+_{\kk_1,+,\s}\psi^-_{\kk_2,+,\s}
\psi^+_{\kk_3,+,\s}\psi^-_{\kk_4-\pp,+,-\s} \r_{\pp,-,-\s}>_T]\}$$
where
$$G^2_\o(\kk)=<\psi^-_{\kk,\o,\s}\psi^+_{\kk,\o,\s}>_T\Eq(s)$$
Similar Dyson equations holds for 
$<\psi^+_{\kk_1,+,\s}\psi^-_{\kk_2,+,\s}\psi^-_{\kk_3,-,\s}\psi^-_{\kk_4,-,\s}>_T$
and $<\psi^+_{\kk_1,+,\s}\psi^-_{\kk_2,+,\s}\psi^-_{\kk_3,+,\s}\psi^-_{\kk_4,+,\s}>_T$.
The Renormalization Group analysis of the preceding sections easily implies
(for details, see [BM3]) that, if $|\bar\kk_i|=\g^h$,$i=1,2,3,4$ 
$$<\psi^+_{\bar\kk_1,+,\s}\psi^-_{\bar\kk_2,+,\s}\psi^+_{\bar\kk_3,-,-\s}
\psi^-_{\bar\kk_4,-,-\s}>_T\equiv G^4_{+,\s}(\bar\kk_1,\bar\kk_2,\bar\kk_3,\bar\kk_4)
=\g^{-4 h}Z_h^{-2}[g^o_{2,h}+O(\bar g_h^2)]\Eq(nbbxc)$$
if $\bar g_h=\sup_{k\ge h}(|g^o_{2,k}|+|g^p_{2,k}|+|g_{4,k}|)$. 
In the Dyson equations appear the functions 
$$G^{2,1}_{\o,\s,\o',\s'}(\kk-\qq,\kk,\qq)=
<\psi^+_{\kk,\o,\s}\psi^-_{\qq,\o,\s} \r_{\kk-\qq,\o',\s'}>_T\Eq(b2)$$
$$G^{4,1}_{+,\s,\o'\s'}(\kk_1,\kk_2,\kk_3,\kk_4-\pp;\pp)=<\psi^+_{\kk_1,+,\s}\psi^-_{\kk_2,+,\s}
\psi^+_{\kk_3,-,-\s}\psi^-_{\kk_4-\pp,-,-\s} \r_{\pp,\o',\s'}>_T\Eq(b2a)$$
Either such functions or the Schwinger functions can be obtained by deriving the {\it Generating functional}
$$\WW(\phi,J)=\log\int P_L(d\psi)e^{-V_L(\psi)+\sum_{\o,\s}\int d\xx (J_{\xx,\o,\s}
\psi^+_{\xx,\o,\s}\psi^-_{\xx,\o,\s}+
\phi^+_{\xx,\o,\s}\psi^-_{\xx,\o,\s}+\psi^+_{\xx,\o,\s}\phi^-_{\xx,\o,\s}
)}\Eq(gfd)$$
with respect to the {\it external fields} $J_{\xx,\o,\s}$ or 
$\phi^-_{\xx,\o,\s}$.

The functions $G^{2,1}, G^{4,1}$ 
are related by remarkable {\it Ward Identities} to the Schwinger functions
$G^2, G^4$. In fact, by operating in \equ(gfd)
the (local) Gauge transformation $\psi^\pm_{+,\s}\to e^{\pm \a_\xx} \psi^\pm_{+,\s}$,
$\psi^\pm_{+,-\s}\to \psi^\pm_{+,-\s}$ and
$\psi^\pm_{-,\pm \s}\to \psi^\pm_{-,\pm \s}$ 
and deriving with respect to $\phi^+_{\yy,+,\s}, \phi^+_{\zz,+,\s}$,
we get, passing to momentum space,
the following Ward Identity
$$D_+(\pp) G^{2,1}_{+,\s,+,\s}(\kk_1-\kk_2,\kk_1,\kk_2)=
G^2_+(\kk_1)-G^2_+(\kk_2)+\D^{2,1}_{+,\s,+,\s}(\kk_1-\kk_2,\kk_1,\kk_2)\Eq(b3)$$
where
$$\D^{2,1}_{\o,\s,\o',\s'}(\kk-\qq,\kk,\qq)=
<\psi^+_{\kk,\o,\s}\psi^-_{\qq,\o,\s} \d\r_{\kk-\qq,\o',\s'}>_T\Eq(b222)$$
$$\d\r_{\kk-\qq,\o,\s}=
\int d\kk' C(\kk',\kk'-\qq)
\psi^+_{\kk',\o,\s}\psi^-_{\kk'-\kk+\qq,\o,\s}\Eq(b2x)$$
$$C(\kk^+,\kk^-)=(C_{h,0}(\kk^-)-1)D_\o(\kk^-)-
(C_{h,0}(\kk^+)-1)D_\o(\kk^+)\Eq(llm)$$
The cutoff function $C_{h,0}$ in $P_L(d\psi)$ destroys the local Gauge invariance
of the theory, and it is responsible of the 
{\it correction term} $\D^{2,1}$ in \equ(b3). 
As explained in
\S 4 of [BM2], 
$\bar\D^{(ij)}\equiv C(\kk,\qq) g^{(i)}(\kk)g^{(j)}(\qq)$
is non vanishing if at least one among $i$ or $j$ is $0$ or $h$; 
this means that either at least one field in $\d\r$ is contracted at scale $0$,
or at least one field in $\d\r$ is contracted at scale $h$. 
We can split the correction term in the following way 
$$\D^{2,1}_{\o,\s,\o',\s'}=\D^{2,1,\a}_{\o,\s,\o',\s'}+\D^{2,1,\b}_{\o,\s,\o',\s'}\Eq(kklbb)$$
where in $\D^{2,1,\a}_{\o,\s,\o',\s'}$ there are all the contributions with
one of the fields
in $\d\r$ contracted at scale $0$, and $\D^{2,1,\b}_{\o,\s,\o',\s'}$
is the rest. It is easy to check that 
$$|\D^{2,1,\b}_{\o,\s,\o',\s'}(\bar\kk_1-\bar\kk_2,\bar\kk_1,\bar\kk_2)|\le 
\bar g_h {\g^{-2h}\over Z_h}\Eq(oom1)$$
This follows from the bound $|\bar\D^{(hj)}|\le \g^{h-j}{\g^{-h-j}\over Z_j}$
and noting that the factor $\g^{h-j}$ gives the correct power counting
for the marginal terms linear in $J$,see [BM2]; note also that the contributions
of order $0$ in the $v_h^L$ cancels out. 

The analysis of $\D^{2,1,\a}_{\o,\s,\o',\s'}$ is more complex;
there are other remarkable identities (first discovered 
in [BM3] for the spinless case) called {\it correction identities} to the 
functions $G^{2,1}$. It holds in fact the following Lemma.
\*
{\cs Lemma 5} {\it There exists functions $\n_{\o,\pm\s}$ such that 
$|\n_{\o,\pm\s}|\le C\bar g_h$ and 
$$\D^{2,1,\a}_{+,\s,+,\s}=
\n_{+,\s}^a D_+(\pp) G^{2,1}_{+,\s,+,\s}$$
$$+\n_{+,-\s}^a 
D_+(\pp) G^{2,1}_{+,\s,+,-\s}+\n_{-,\s}^a D_-(\pp) G^{2,1}_{+,\s,-,\s}
+\n_{-,-\s}^a D_-(\pp) G^{2,1}_{+,\s,-,-\s}+H^{2,1,\a}_{+,\s,+,\s}
\Eq(b7)$$
with,if $|\bar\kk_1|=|\bar\kk_2|=\g^h$
$$|H^{2,1,\a}_{+,\s,+,\s}(\bar\kk_1-\bar\kk_2,\bar\kk_1,\bar\kk_2)|
\le C {\g^{-2 h}\over Z_h^2}\g^{\th h}\Eq(bhhg)$$
for some constants $C$ and $0<\th<1$.}
\*
The above identity says that the correction $\D^{2,1,\a}$ can be written in terms of 
the functions $G^{2,1}$, up to a term which is smaller than $O(\g^{\th h})$. 
We will call $H^{2,1}_{a}=H^{2,1,\a}_{+,\s,+,\s}+\D^{2,1,\b}_{+,\s,+,\s}$, and
$$|H^{2,1}_{a}(\bar\kk_1-\bar\kk_2,\bar\kk_1,\bar\kk_2)|\le 
\bar g_h {\g^{-2h}\over Z_h}\;.\Eq(oom)$$
By the phase
transformation  $\psi^\pm_{+,-\s}\to e^{\pm \a_\xx} \psi^\pm_{+,-\s}$,
$\psi^\pm_{+,\s}\to \psi^\pm_{+,\s}$ and
$\psi^\pm_{-,\pm \s}\to \psi^\pm_{-,\pm \s}$, and using
a correction identity similar to \equ(b7) we find
$$-\n_{+,\s}^b D_+(\pp) G^{2,1}_{+,\s,+,\s}+(1-\n_{+,-\s}^b) 
D_+(\pp) G^{2,1}_{+,\s,+,-\s}-\n_{-,\s}^b D_-(\pp) G^{2,1}_{+,\s,-,\s}
-\n_{-,-\s}^b D_-(\pp) G^{2,1}_{+,\s,-,-\s}=H^{2,1}_{b}\Eq(b61)$$
where $H^{2,1,a}_{b}$ verifies a bound similar to \equ(oom).

In the same way by
the Gauge transformation $\psi^\pm_{-,\s}\to e^{\pm \a_\xx} \psi^\pm_{-,\s}$,
$\psi^\pm_{-,-\s}\to \psi^\pm_{-,-\s}$ and
$\psi^\pm_{+,\pm \s}\to \psi^\pm_{+,\pm \s}$ , and using
a correction identity similar to \equ(b7) we find
$$-\n_{+,\s}^c D_+(\pp) G^{2,1}_{+,\s,+,\s}-\n_{+,\s}^c 
D_+(\pp) G^{2,1}_{+,\s,+,-\s}+(1-\n_{-,\s}^c) D_-(\pp) G^{2,1}_{+,\s,-,\s}
-\n_{-,-\s}^c D_-(\pp) G^{2,1}_{+,\s,-,-\s}=H^{2,1}_{c}\Eq(b62)$$
where $H^{2,1,a}_{c}$ verifies a bound similar to \equ(oom).

Finally by
the Gauge transformation $\psi^\pm_{-,-\s}\to e^{\pm \a_\xx} \psi^\pm_{-,-\s}$,
$\psi^\pm_{-,\s}\to \psi^\pm_{-,\s}$ and
$\psi^\pm_{+,\pm \s}\to \psi^\pm_{+,\pm \s}$ we get the following WI
$$-\n_{+,\s}^d D_+(\pp) G^{2,1}_{+,\s,+.\s}-\n_{+,-\s}^d 
D_+(\pp) G^{2,1}_{+,\s,+,-\s}-\n_{-,\s}^d D_-(\pp) G^{2,1}_{+,\s,-,\s}
+(1-\n_{-,-\s}^d) D_-(\pp) G^{2,1}_{+,\s,-,-\s}=H^{2,1}_{d}\Eq(b63)$$
Bounds like \equ(oom) and \equ(bhhg) hold also for 
$H^{2,1}_{b}, H^{2,1}_{c},H^{2,1}_{d}$.
It is easy to see from some algebra that the above relations imply 
$$D_+(\pp) G^{2,1}_{+,\s,+,\s}(\kk_1-\kk_2,\kk_1,\kk_2)=
G^2_+(\kk_1)-G^2_+(\kk_2)+(1+F_a^1)H^{2,1}_a+F_b^1 H^{2,1}_b+
F_c^1 H^{2,1}_c+F_d^1 H^{2,1}_d
\Eq(b33)$$
$$D_+(\pp) G^{2,1}_{+,\s,+,-\s}(\kk_1-\kk_2,\kk_1,\kk_2)=
(1+F_a^2)H^{2,a}_a+F_b^2 H^{2,1}_b+
F_c^2 H^{2,1}_c+F_d^2 H^{2,1}_d
\Eq(b34)$$
$$D_-(\pp) G^{2,1}_{+,\s,-,\s}(\kk_1-\kk_2,\kk_1,\kk_2)=
(1+F_a^3)H^{2,a}_a+F_b^3 H^{2,1}_b+
F_c^3 H^{2,1}_c+F_d^3 H^{2,1}_d
\Eq(b34)$$
$$D_-(\pp) G^{2,1}_{+,\s,-,-\s}(\kk_1-\kk_2,\kk_1,\kk_2)=
(1+F_a^4)H^{2,a}_a+F_b^4 H^{2,1}_b+
F_c^4 H^{2,1}_c+F_d^4 H^{2,1}_d
\Eq(b34)$$
with $F_a,F_b,F_c,F_d$ are combinations of the $\n$, with the property
that if $|\n_i^j|\le C\bar g$, then $|F_i|\le C\bar g_h$.
Then \equ(b33)
really provides a relation between $G^{2,1}$ and $G^2$ up to bounded
corrections.
\vskip.5cm
\* \sub(5.2b)
{\it Proof of Lemma 5.} 

We introduce 
the generating function for $H^{2,1}_a$
$$\int P_L(d\psi) e^{-\VV_L+T_1+\int d\kk  d\pp J_\pp \nu_{+,\s}^a
D_+(\pp)\psi^+_{\kk,+,\s}\psi^-_{\kk+\pp,+\s}}\Eq(llmn)$$
$$e^{-\int d\kk  d\pp J_\pp[
\nu_{+,-\s}^a
D_+(\pp)\psi^+_{\kk,+,-\s}\psi^-_{\kk+\pp,+,-\s}+\nu_{-,\s}^a
D_-(\pp)\psi^+_{\kk,-,\s}\psi^-_{\kk+\pp,-,\s}+
\nu_{-,-\s}^a D_-(\pp)\psi^+_{\kk,-,-\s}\psi^-_{\kk+\pp,-,-\s}]}$$
where
$$T_1=\int d\kk d\pp J_\pp 
C(\kk,\kk+\pp)\psi^+_{\kk,+,\s}\psi^-_{\kk+\pp,+,\s}\Eq(llkmm)$$
The analysis proceeds essentially identical to the one of 
\S 4 of [BM2]. After integrating the $\psi^0$ field, we get
in the effective potential a sum of monomials  
of the form $W J^m \psi_1...\psi_n$; we extend the definition
of $\LL$ to monomials of this kind by requiring
that it acts non trivially only on the terms linear in $J$
and quadratic in $\psi$, as a power counting argument shows that they
are the only marginal terms.

Consider now the terms in which $T_1$ is contracted;
they are of the form
$$
\sum_{\tilde \o,\tilde\s} \int d\pp \int d\kk^+ J_\pp
\hat\psi^+_{\kk^+,\tilde\o,\tilde\s} \hat\psi^-_{\kk^+ -\pp,\tilde\o,\tilde\s} 
[ F^{(-1)}_{2,+,\s,\tilde\o,\tilde\s}(\kk^+, \kk^+-\pp) +
F^{(-1)}_{1,+,\s}(\kk^+, \kk^+-\pp) \d_{+,\tilde\o}\d_{\s,\tilde\s}]\Eq(3.28)$$
where $F^{(-1)}_{2,+,\s,\tilde\o,\tilde\s}$ 
is given by all the terms obtained contracting both the $\psi$ fields 
in $T_1$ while $F^{(-1)}_{1,+,\s}$ are is given by the terms obtained
leaving external one of the $\psi$-fields of $T_1$
Both contributions to the r.h.s. of \equ(3.28) are
dimensionally marginal; however, the renormalization of
$F^{(-1)}_{1,\o,\s}$ is trivial, as it is of the form
$$F_{1,+,\s}^{(-1)}(\kk^+,\kk^-)=
[{[C_{h,0}(\kk^-)-1] D_+(\kk^-) \hat g^{(0)}_\o(\kk^+)- u_0(\kk^+)
\over D_+(\kk^+-\kk^-)} G^{(2)}(\kk^+)\Eq(3.28a)$$
or the similar one, obtained exchanging $\kk^+$ with $\kk^-$.

By the oddness of the propagator in the momentum,
$G^{(2)}(0)=0$, hence we can regularize  such term without
introducing any local term, by simply rewriting it as
$$F_{1,+,\s}^{(-1)}(\kk^+,\kk^-)=
[{[C_{h,0}(\kk^-)-1] D_\o(\kk^-) \hat g^{(0)}_+(\kk^+)- u_0(\kk^+)
\over D_\o(\kk^+-\kk^-)} [G^{(2)}(\kk^+)-G^{(2)}(0)]\;.\Eq(3.28b)$$

As shown in [BM2], by using the symmetry property
$$\hat g^{(j)}_\o(\kk)=-i\o \hat g^{(j)}_\o(\kk^*) \virg
\kk=(k,k_0),\quad \kk^*=(-k_0,k)\;,\Eq(3.28c)$$
$F^{(-1)}_{2,\o,\s,\tilde\o,\tilde\s}$ can be written as
$$F^{-1}_{2,\o,\s,\tilde\o,\tilde\s}(\kk^+, \kk^-) = {1\over D_\o(\pp)}
\left[ p_0 A_{0,\o,\s,\tilde\o,\tilde\s}(\kk^+,\kk^-) + p_1
A_{1,\o,\s,\tilde\o,\tilde\s}(\kk^+,\kk^-) \right]\;,\Eq(3.29a)$$
where $A_{i,\o,\s,\tilde\o,\tilde\s}(\kk^+,\kk^-)$ are functions such that, if
we define
$$\LL F^{-1}_{2,+,\s,\tilde\o,\pm \s}={1\over D_+(\pp)}
\left[ p_0 A_{0,+,\s,\tilde\o,\pm \s}(0,0) +p_1 A_{1,+,\s,\tilde\o,\pm \s}(0,0)
\right]\;,\Eq(3.29)$$
then,
$$\LL F^{-1}_{2,+,\s,\tilde\o,\pm\s}= D_{\tilde\o}(\pp) Z_{-1}^{3,\tilde\o,\pm\s}\;,\Eq(3.30)$$
where  $Z_{-1}^{3,\tilde\o,\pm\s}$ are four suitable real
constants. 

Consider now the terms in which the $\n_{\o,\s}$ are contracted;
we define the localization operator on such terms as
$$\LL \int d\kk d\pp D_\o(\pp)W^{-1}_{\o,\pm\s}(\kk,\kk-\pp)
\psi^+_{\kk,\o,\pm\s}\psi^-_{\kk+\pp,\o,\pm\s}=
\int d\kk d\pp D_\o(\pp)W^{-1}_{\o,\pm\s}({\bf 0},{\bf 0})
\psi^+_{\kk,\o,\pm\s}\psi^-_{\kk+\pp,\o,\pm\s}
\Eq(llccv)$$
We define $\n_{-1,\o,\pm\s}=Z_{-1}^{3,\o,\pm\s}+W^{-1}_{\o,\pm\s}$
we get that
the local terms linear in $J$ are
$$\int d\kk  d\pp J_\pp[\nu_{-1,+,\s}^a
D_+(\pp)\psi^+_{\kk,+,\s}\psi^-_{\kk+\pp,+\s}+\Eq(peff)$$
$$\nu_{-1,+,-\s}^a
D_+(\pp)\psi^+_{\kk,+,-\s}\psi^-_{\kk+\pp,+,-\s}+\nu_{-1,-,\s}^a
D_-(\pp)\psi^+_{\kk,-,\s}\psi^-_{\kk+\pp,-,\s}+
\nu_{-1,-,-\s}^a D_-(\pp)\psi^+_{\kk,-,-\s}\psi^-_{\kk+\pp,-,-\s}]$$
We can iterate the above procedure; at the integration
of the generic scale the terms quadratic and linear in $J$ 
in the effective potential are obtained contracted a $T_1$ vertex (in such a case
one of the two fields of $T_1$ is necessarily contracted at scale $0$)
or a $\n_{k,\o,\s}$ vertex; in both case the preceding analysis can be repeated and 
the local terms linear in $J$ are, for $k> h$
$$\int d\kk  d\pp J_\pp[\nu_{k,+,\s}^a
D_+(\pp)\psi^+_{\kk,+,\s}\psi^-_{\kk+\pp,+\s}+\Eq(peff)$$
$$\nu_{k,+,-\s}^a
D_+(\pp)\psi^+_{\kk,+,-\s}\psi^-_{\kk+\pp,+,-\s}+\nu_{k,-,\s}^a
D_-(\pp)\psi^+_{\kk,-,\s}\psi^-_{\kk+\pp,-,\s}+
\nu_{k,-,-\s}^a D_-(\pp)\psi^+_{\kk,-,-\s}\psi^-_{\kk+\pp,-,-\s}]$$
We have then obtained an expansion 
for $H^{2,1}_a$ in which new running coupling constants appear,namely
$\nu_{k,\o,\pm\s}$; the analogue of Theorem 3 ensures convergence
$\nu_{k,\o,\pm\s}$ are small or any $k>h$. The beta function for 
$\nu_{k,\o,\pm\s}$ has the following form
$$\nu_{k-1,\o,\pm\s}=\nu_{k,+,\pm\s}
+\b^{1,k}_{\o,\pm\s}(v_k^L,...,v_0^L)+\b^{2,k}_{\o,\pm\s}(v_k^L,\n_k...,v_0^L,\n_0)
\Eq(b144)$$
where by definition $\b^{2,k}_{\o,\pm\s}(v_k^L,\n_k...,v_0^L,\n_0)$
is obtained contracting a $\n_j$ while 
$\b^{1,k}_{\o,\pm\s}(v_k^L,...,v_0^L)$ is obtained contracting $T_1$ 
and 
$$|\b^{1,k}_{\o,\pm\s}(v_k^L,...,v_0^L)|\le C\bar g_k\g^{\th k}$$
for some constant $0<\th<1$. The presence of the factor 
$\g^{\th k}$ in the above bound is due to the fact that, for the support
properties of the function $C(\kk^+,\kk^-)$ discussed after \equ(llkmm),
one of the fields of $T_1$ is necessarily contracted at scale $0$.

In fact we can show (proceeding as in the proof of Lemma 3,or in \S 4.6 of BM3)
that there exists a sequence $\n_k$ such that $|\nu_{k,\o,\pm\s}|\le C\bar g_h
\g^{\th k}$ by solving
$$\nu_{k,\o,\pm\s}=-\sum_{k'=h+1}^k\{
\b^{1,k'}_{\o,\pm\s}(v_k^L,...,v_0^L)+\b^{2,k'}_{\o,\pm\s}(v_k^L,\n_k...,v_0^L,\n_0)\}
\Eq(b144b)$$
This shows that there exist $\nu_{\o,\pm\s}$ such that 
$\nu_{k,\o,\pm\s}=O(\g^{\th k})$.

We have then find an expansion for $H^{2,1}_a(\bar\kk_1-\bar\kk_2,\bar\kk_1,\bar\kk_2)$ 
very similar to the one of 
$G^{2,1}$, but in which each tree contributing to 
$H^{2,1}_a(\bar\kk_1-\bar\kk_2,\bar\kk_1,\bar\kk_2)$ 
have an extra $\g^{\th h}$; in fact or there is an endpoint
$\n_k$ (and we use that $\nu_{k,\o,\pm\s}=O(\g^{\th k})$
and the fact that, as the dimension are negative,the value of the tree
has an extra $\g^{\th (h-k)}$) or there is an endpoint $T_1$ contracted at scale $0$
(hence, as the dimensions are negative,the value of the tree
has an extra $\g^{\th h}$).
\qed
\vskip.5cm
Inserting \equ(b34) in the Dyson equation, and using \equ(oom) and \equ(bhhg), 
we see that the first three addends
of the Dyson equation are given by $(g_2^0+O(\bar g_h^2)){\g^{-4 h}\over Z_h^2}$.
\vskip.5cm
We have to consider now the last three addenda in the Dyson equation; let us start by
$$\int d\pp [g_2^o <\psi^+_{\bar\kk_1,+,\s}\psi^-_{\bar\kk_2,+,\s}
\psi^+_{\bar\kk_3,-,-\s}\psi^-_{\bar\kk_4-\pp,-,-\s} \r_{\pp,+,\s}>_T\Eq(last)$$
Let us call
$$G^{4,1}_{+,\s,\o'\s'}(\kk_1,\kk_2,\kk_3,\kk_4-\pp;\pp)=<\psi^+_{\kk_1,+,\s}\psi^-_{\kk_2,+,\s}
\psi^+_{\kk_3,-,-\s}\psi^-_{\kk_4-\pp,-,-\s} \r_{\pp,\o',\s'}>_T\Eq(last1)$$
As $|\bar\kk_4|= \g^h$ the support
properties of the propagators imply that $|\pp|\le \g+\g^h\le
2\g$, hence we can freely multiply $G^{4,1}_+$ in the r.h.s. of
\equ(last) by the compact support function $\c_0(\g^{-j_m}|\pp|)$,
with $j_m= [1+\log_\g 2] + 1$. 
It follows that \equ(last) can be written as
$$\int d\pp \chi_M(\pp)
G^{4,1}_+(\pp;\kk_1,\kk_2,\kk_3,\kk_4-\pp)+ \int d\pp
\tilde\chi_M(\pp) G^{4,1}_+(\pp;\kk_1,\kk_2,\kk_3,\kk_4-\pp)
\Eq(last3)$$
where $\chi_M(\pp)$ is a compact support function vanishing for
$|\pp|\ge \g^{h+j_m-1}$ and
$$\tilde\chi_M(\pp) = \sum_{h_p=h+j_m}^{j_m} f_{h_p}(\pp)\;.\Eq(2.11b)$$
Note that the decomposition of the $\pp$ sum is done so that
$\tilde\chi_M(\pp)=0$ if $|\pp|\le 2\g^h$.
It is easy to show that the first term in \equ(last3) is bounded by
$O({\bar g_h^2 \g^{-3 h}\over Z_h^2})$, see [BM3].
Regarding the second addend we will use
the following Ward identities 
$$(1-\n_{+,\s}^a)D_+(\pp) G^{4,1}_{+,\s,+,\s}
-\n_{+,-\s}^a 
D_+(\pp) G^{4,1}_{+,\s,+,-\s}-\n_{-,\s}^a D_-(\pp) G^{4,1}_{+,\s,-,\s}
-\n_{-,-\s}^a D_-(\pp) G^{4,1}_{+,\s,-,-\s}=$$
$$G^4_+(\kk_1-\pp,\kk_2,\kk_3,\kk_4-\pp)-G^4_+(\kk_1,\kk_2+\pp,\kk_3,\kk_4-\pp)
+H^{4,1}_{a}(\kk_1,\kk_2,\kk_3,\kk_4-\pp;\pp)\Eq(b6)$$
$$-\n_{+,\s}^b D_+(\pp) G^{4,1}_{+,\s,+,\s}+(1-\n_{+,-\s}^b) 
D_+(\pp) G^{4,1}_{+,\s,+,-\s}-\n_{-,\s}^b D_-(\pp) G^{2,1}_{+,\s,-,\s}
-\n_{-,-\s}^b D_-(\pp) G^{4,1}_{+,\s,-,-\s}=H^{4,1}_{b}\Eq(b610)$$
$$-\n_{+,\s}^c D_+(\pp) G^{4,1}_{+,\s,+.\s}-\n_{+,-\s}^c 
D_+(\pp) G^{4,1}_{+,\s,+,-\s}+(1-\n_{-,\s}^c) D_-(\pp) G^{4,1}_{+,\s,-,\s}
-\n_{-,-\s}^c D_-(\pp) G^{4,1}_{+,\s,-,-\s}=H^{4,1}_{c}\Eq(b620)$$
$$-\n_{+,\s}^d D_+(\pp) G^{4,1}_{+,\s,+.\s}-\n_{+,-\s}^d 
D_+(\pp) G^{4,1}_{+,\s,+,-\s}-\n_{-,\s}^d D_-(\pp) G^{2,1}_{+,\s,-,\s}
+(1-\n_{-,-\s}^d) D_-(\pp) G^{4,1}_{+,\s,-,-\s}=$$
$$G^4_+(\kk_1,\kk_2,\kk_3-\pp,\kk_4-\pp)-G^4(\kk_1,\kk_2,\kk_3,\kk_4)+H^{2,1}_{d}
(\kk_1,\kk_2,\kk_3,\kk_4-\pp;\pp)\Eq(b630)$$
where the functions $H^{4,1}_i$ are defined in an analogous way to the 
functions $H^{2,1}_i$.
It is easy to see from some algebra that the above relations imply 
$$D_+(\pp) G^{4,1}_{+,\s,-,-\s}(\kk_1,\kk_2,\kk_3,\kk_4-\pp;\pp)=(1+G_1^a)
[G^4_+(\kk_1-\pp,\kk_2,\kk_3,\kk-\pp)-G^4_+(\kk_1-\pp,\kk_2,\kk_3,\kk-\pp)]$$
$$+G_2^a [G^4_+(\kk_1,\kk_2,\kk_3-\pp,\kk_4-\pp)-G^4(\kk_1,\kk_2,\kk_3,\kk_4)]
+(1+G_3^a)H^{4,a}_a+G_4^a H^{4,1}_b+
G^5_a H^{4,1}_c+G^6_a H^{4,1}_d
\Eq(b33)$$
with $G^i_a=O(\bar g)$. From \equ(b33) 
we can decompose $\int d\pp
\tilde\chi_M(\pp) G^{4,1}_+$ as sum of several terms; the one involving 
$G^4(\kk_1,\kk_2,\kk_3,\kk_4)$ is vanishing while the other three terms involving
the other functions $G^4$ have a bound 
$O({\bar g_h^2 \g^{-3 h}\over Z_h^2})$, see [BM3].
Finally the following results holds 
%
%
\*
{\cs Lemma 6} {\it If the functions
$\n_{\o,\pm\s}$ are the same as in Lemma 7,it holds that, for $i=a,b,c,d$
$$|\int d\pp g_-(\kk_4) {H^{4,1}_i\over D_+(\pp)}|\le C 
{\bar g_h^2 \g^{-3 h}\over Z_h^2}\Eq(fon1)$$
}
\*
Inserting all the above bounds in the Dyson equation \equ(b1)
computed at momenta $|\kk_i|=\g^h$,$i=1,2,3,4$ we have completed
the proof of Lemma 3.
\* 
Lemma 6 is proved considering
$$\tilde G^4_+(\kk_1, \kk_2,\kk_3, \kk_4) = \left.
{\partial^4\over \partial \phi^+_{\kk_1,+,\s} \partial
\phi^-_{\kk_2,+,\s} \partial \phi^+_{\kk_3,-,-\s}
\partial J_{\kk_4}} \tilde W \right|_{\phi=0}\;,\Eq(3.2)$$
where
$$\tilde W = \log \int P(d\hat\psi)e^{-T(\psi) + \n_1^a T_1(\psi)
+\n_2^a T_2(\psi)+\n_3^a T_3(\psi)+\n_4^a T_4(\psi)
+ \sum_\o\int d\xx
[\phi^+_{\xx,\o}\hat\psi^{-}_{\xx,\o}+
\hat\psi^{+}_{\xx,\o}\phi^-_{\xx,\o}]} \;,\Eq(3.3)$$
$$T(\psi) = {1\over L\b} \sum_{\pp} \tilde\chi_M(\pp)
{1\over L\b} \sum_{\kk} {C_+(\kk, \kk-\pp) \over D_+(\pp)}
(\hat\psi_{\kk,+,\s}^+ \hat\psi_{\kk-\pp,+,\s}^-)
\hat\psi^+_{\kk_4-\pp,-,-\s} \hat J_{\kk_4} \hat
g_-(\kk_4)\;,\Eq(3.4)$$
$$T_1(\psi)= {1\over L\b} \sum_{\pp} \tilde\chi_M(\pp)
{1\over L\b} \sum_{\kk} (\hat\psi_{\kk,+,\s}^+
\hat\psi_{\kk-\pp,+,\s}^-) \hat\psi^+_{\kk_4-\pp,-,-\s} \hat J_{\kk_4}
\hat g_-(\kk_4)\;,\Eq(3.5v)$$
$$T_2(\psi)= {1\over L\b} \sum_{\pp} \tilde\chi_M(\pp)
{1\over L\b} \sum_{\kk} (\hat\psi_{\kk,+,-\s}^+
\hat\psi_{\kk-\pp,+,-\s}^-) \hat\psi^+_{\kk_4-\pp,-,-\s} \hat J_{\kk_4}
\hat g_-(\kk_4)\;,\Eq(3.5vbvg)$$
$$T_3(\psi) = {1\over L\b} \sum_{\pp} \tilde\chi_M(\pp)
{1\over L\b} \sum_{\kk} {D_-(\pp) \over D_+(\pp)}
(\hat\psi_{\kk,-,\s}^+ \hat\psi_{\kk-\pp,-,\s}^-)
\hat\psi^+_{\kk_4-\pp,-,-\s} \hat J_{\kk_4} \hat
g_-(\kk_4)\;.\Eq(3.6n)$$
$$T_4(\psi) = {1\over L\b} \sum_{\pp} \tilde\chi_M(\pp)
{1\over L\b} \sum_{\kk} {D_-(\pp) \over D_+(\pp)}
(\hat\psi_{\kk,-,-\s}^+ \hat\psi_{\kk-\pp,-,-\s}^-)
\hat\psi^+_{\kk_4-\pp,-,-\s} \hat J_{\kk_4} \hat
g_-(\kk_4)\;.\Eq(3.6n)$$
It holds that 
$$\tilde G^4_+(\bar\kk_1,\bar\kk_2,\bar\kk_3,\bar\kk_4)
=\int d\pp g_-(\kk_4) {H^{4,1}_i\over D_+(\pp)}\Eq(kkkl)$$
Note that the expansion of $\tilde G_+^4$
is very similar to the expansion of $G_+^4$, except for the presence of a special vertex associated to $J$.
The proof of the bound \equ(fon1) 
is essentially identical to the one for the spinless
case of [BM3], to which we refer for the technical details.
\vskip.5cm
\section(7, Correlation functions)
\vskip.5cm
%
%
Once that the multiscale analysis
of the partition function is completed,
it is possible to apply the same ideas and methods to the 
Grassmann integrals giving the Schwinger function or 
the correlations; as the analysis is essentially identical
to the one in [BM1], we will give only the main ideas referring
to \S 5 of [BM1] for details.

The density-density correlation can be written in terms of a Grassmann integral
in the following way
$$<\r(\xx)\r(\yy)>_T=<\r(\xx)\r(\yy)>-<\r(\xx)><\r(\yy)>
={\partial^2 \SS\over\partial\phi(\xx)\partial\phi(\yy)}\Eq(parp)$$
where
$$\SS(\phi)=\log\int P(d\psi)e^{-\VV-\sum_\s\int d\xx\phi(\xx)
\psi^{+}_{\xx,\s} \psi^{-}_{\xx,\s}}\Eq(bbllk)$$
We shall evaluate $\SS$ in a way which
is very close to that used for the integration of the partition function
in \S 2. We introduce
the scale decomposition described above and we perform iteratively
the integration of the single scale fields, starting from the field of
scale $1$. 
 
After integrating the fields $\psi^{(1)},...\psi^{(h+1)}$
we find
$$e^{\SS(\phi)}=e^{-L\b E_h+S^{(h+1)}(\phi)}\int P_{Z_h,C_h}(d\psi^{\le
h})e^{-\VV^{(h)}(\sqrt{Z_h}\psi^{(\le h)})+\BB^{(h)}
(\sqrt{Z_h}\psi^{(\le h)},\phi)}\;,\Eq(6.4)$$
where $P_{Z_h,\s_h,C_h}(d\psi^{(\le h)})$ and $\VV^{h}$ are given by 
\equ(4.4a) and \equ(2.61a), respectively, while $S^{(h+1)}$ $(\phi)$, which
denotes the sum over all the terms dependent on $\phi$ but independent of
the $\psi$ field, and $\BB^{(h)}(\psi^{(\le h)}, \phi)$, which denotes the
sum over all the terms containing at least one $\phi$ field and two $\psi$
fields, can be represented in the form, if $\int d\xx=\int_{-{\b\over 2}}^{\b\over 2} 
dx_0\sum_{x\in\L}$
$$S^{(h+1)}(\phi)=\sum_{m=1}^\io\int d\xx_1\cdots d\xx_m
S^{(h+1)}_m(\xx_1,\ldots,\xx_m)
\Big[\prod_{i=1}^m\phi(\xx_i)\Big]\Eq(6.5)$$
$$\eqalign{
&\BB^{(h)}(\psi^{(\le h)},\phi)=\sum_{m=1}^\io\sum_{n=1}^{\io} \sum_{\ss,\oo}
\int d\xx_1\cdots d\xx_m d\yy_1 \cdots d\yy_{2n} \;\cdot\cr
&\qquad\cdot\; B^{(h)}_{m,2n,\ss,\oo}(\xx_1,\ldots,\xx_m;\yy_1,\ldots,\yy_{2n})
\Big[\prod_{i=1}^m\phi(\xx_i)\Big] \Big[\prod_{i=1}^{2n}
\psi^{(\le h)\s_i}_{\yy_i,\o_i}\Big]\;.\cr}\Eq(6.6)$$
 
Since the field $\phi$ is equivalent, from the point of view of dimensional
considerations, to two $\psi$ fields, the only terms in the r.h.s. of
\equ(6.6) which are not irrelevant are those with $m=1$ and $n=1$, which are
marginal. 
Hence we extend the definition of the
localization operator $\LL$, so that its action on $\BB^{(h)}(\psi^{(\le
h)},\phi)$ in described in the following way, by its action on the kernels
$B^{(h)}_{m,2n,\s,\o}(\pp,\kk_1,..,\kk_n)$:
 
\*
\0 1) if $m=1$, $n=1$ then
$$\LL B^{(h)}_{1,2,\s,\o}(\pp;\kk_1,\kk_2)=
B^{(h)}_{1,2,\s,\o}(0;0,0)\Eq(6.7bb)$$
 
\0 2) $\LL=0$ in all the other cases

It follows that
$$\LL \BB^{(h)}(\psi^{(\le h)},\phi)={Z^{(1)}_h\over Z_h} F_1^{(\le h)}
+{Z^{(2)}_h\over Z_h} F_2^{(\le h)}\;,\Eq(6.1ll2)$$
where $Z^{(1)}_h$ and $Z^{(2)}_h$ are real numbers, such that
$Z^{(1)}_1=Z^{(2)}_1=1$ and
$$F_1^{(\le h)}=\sum_{\o,\s}\int d\xx \phi(\xx)e^{2i\o \pp_F x}
\psi^{(\le h)+}_{\xx,\o,\s}\psi^{(\le h)-}_{\xx,-\o,\s}\;,\Eq(6.13cc)$$
$$F_2^{(\le h)}=\sum_{\s=\pm 1}\int d\xx \phi(\xx)
\psi^{(\le h)\s}_{\xx,\o,\s}\psi^{(\le h)-}_{\xx,\o,\s}\;.\Eq(6.14cc)$$
 
By using the notation of \S 2, we can write the integral in the r.h.s.
of \equ(6.4) as
$$\eqalign{
&e^{-L\b t_h} \int P_{\tilde Z_{h-1},C_h}(d\psi^{(\le h)})
e^{-\tilde\VV^{(h)}(\sqrt{Z_h}\psi^{(\le h)})+\BB^{(h)}
(\sqrt{Z_h}\psi^{(\le h)},\phi)}\;=\cr
&= e^{-L\b t_h} \int P_{Z_{h-1},C_{h-1}}(d\psi^{(\le h-1)})\;\cdot\cr
&\cdot\; \int P_{Z_{h-1},\tilde f_h^{-1}}(d\psi^{(h)})
e^{-\hat\VV^{(h)}(\sqrt{Z_{h-1}}\psi^{(\le h)})+\hat\BB^{(h)}
(\sqrt{Z_{h-1}}\psi^{(\le h)},\phi)}\;,\cr}\Eq(6.15)$$
where $\hat\VV^{(h)}(\sqrt{Z_{h-1}}\psi^{(\le h)})$ is defined as in \S 3 and
$$\hat\BB^{(h)}(\sqrt{Z_{h-1}}\psi^{(\le h)},\phi)=
\BB^{(h)}(\sqrt{Z_{h}}\psi^{(\le h)},\phi)\;.\Eq(6.16vv)$$
$\BB^{(h-1)}(\sqrt{Z_{h-1}}\psi^{(\le h-1)},\phi)$ and $S^{(h)}(\phi)$
are then defined through 
$$\eqalign{
&e^{-\VV^{(h-1)}(\sqrt{Z_{h-1}}\psi^{(\le h-1)})+\BB^{(h-1)}
(\sqrt{Z_{h-1}}\psi^{(\le h-1)},\phi)-L\b\tilde E_h+\tilde S^{(h)}(\phi)}=\cr
&=\int P_{Z_{h-1},\tilde f_h^{-1}}(d\psi^{(h)})
e^{-\hat\VV^{(h)}(\sqrt{Z_{h-1}}\psi^{(\le h)})+\hat\BB^{(h)}
(\sqrt{Z_{h-1}}\psi^{(\le h)},\phi)}\;.\cr}\Eq(6.17)$$
 
Of course also the new renormalization constants related
to the density-density correlation function obey to a Beta function
equation of the form  
$${Z^{(i)}_{h-1}\over Z^{(i)}_h} = 1 + z^{(i)}_h\;,\quad i=1,2\;,\Eq(6.18)$$
where $z^{(1)}_h$ and $z^{(2)}_h$ are some quantities of order $\bar g_h$.
It turns out that $\lim_{h\to-\io}{Z_h^1\over \g^{\h_1 h}}=1+O(U)$
while $\lim_{h\to-\io}Z_h^1=1+O(U)$, with $\h_1=- b U+O(U^2)$ and $b>0$
is a suitable constant.
%
%
%
%
The bounds for the expansion of the Schwinger function
or the correlation functions are done exactly as in \S 5 of
[BM1]; to the first term in
\equ(1.3aa) or to the first two terms in \equ(1.12) contribute only trees
with only endpoints with scale $\le 0$; the other 
trees have at least an endpoint at scale $1$ so that 
by the short memory property
they have a faster decay. 
\vskip.5cm
\section(5SSX, The Hubbard model in a magnetic field)
\vskip.5cm
We only sketch the analysis when there is a magnetic field
as it is indeed very similar to analysis of the vanishing magnetic field case.

The presence of a magnetic field destroys the SU(2) spin symmetry.
The counterterms are introduced by the following definition
$$\tilde t=t-\sum_\s \d_\s\quad \cos p_F^\s=\m+\sign(\s) h-\n_\s\Eq(ccd1)$$
This means that $\VV$ in the partition function \equ(h1) is replaced by 
$$\VV=U \int_{-\b/2}^{\b/2} dx_0 \sum_x 
\psi^+_{\xx,+}\psi^-_{\xx,+}\psi^+_{\xx,-}\psi^-_{\xx,-}+ 
\int_{-\b/2}^{\b/2} d x_0 \sum_{x,\s}\n_\s
\psi^+_{\xx,\s}\psi^-_{\xx,\s}+ 
\int_{-\b/2}^{\b/2} d x_0 \sum_{x,y,\s}\d_\s t_{x,y} 
\psi^+_{\xx,\s}\psi^-_{\xx,\s}\Eq(v)$$

The ultraviolet and infrared integration are done as in \S 2,\S 3,
with the difference that for $h\le\bar h$ only quartic momomials verifying 
$||\sum_{i=1}^4 \e_i \o_i p_F^{\s_i}||=0$ (instead of \equ(cond))
are present in the effective potential. The definition of $\LL$
on the quartic terms is similar to \equ(1.18) with the difference that the delta function
in the \equ(1.18) is replaced by $\d(\sum_{i=1}^4 \e_i \o_i p_F^{\s_i})$.
This means that the quartic marginal terms verify
$\sum_{i=1}^4 \e_i \o_i p_F^{\s_i}=0$ mod. $2\pi$
and this condition forbids the configuration of 
$\o$ given by the second of \equ(poss), if $h$ is small enough,
as $p_F(\s)-p_F(-\s)+ n\pi\not=0$; in other words there is no
the analogue of the $g_1^p$-terms in the effective potential. 
Moreover we are assuming in Theorem 2 that $|\cos^{-1}(\m+h)+\cos^{-1}(\m-h)-\pi|\ge \bar C$
for some constant $C$; 
this implies that the configuration $(\o,-\o,\o,-\o)$
is not allowed.

The wave function renormalization depends by $\s$, that is $Z_{h,\s})$
and the relevant part of the effective potential is given by
$$\LL{\widehat\VV}^{(h)}(\sqrt{Z_{h-1}}\psi^{(\le h)})=
\sum_\s \{\d_{h,\s} F_{a,\s}(\sqrt{Z_{h-1}}\psi^{(\le h)}+
\g^h\n_{h,\s} F_{n,\s}(\sqrt{Z_{h-1}}\psi^{(\le h)})\Eq(nnbllp)$$
$$+g_{2,h,\s}^p F_{2,\s,-\s}(\sqrt{Z_{h-1}}\psi^{(\le h)}+g_{2,h,\s}^o F_{2,\s,\s}(\sqrt{Z_{h-1}}\psi^{(\le h)}+
g_{4,h,\s} F_{4,\s,-\s}(\sqrt{Z_{h-1}}\psi^{(\le h)}$$
There are then in the $h\not=0$ case $4$ quadratic and $6$
quartic running coupling constants; Theorem 3 is still valid
if they are small enough.
We  can choose $\n_\s,\d_\s$ so that 
$\n_{h,+},\n_{h,-}, \d_{h,+},\d_{h,-}$ are 
$O(\bar g_h\g^{\th h})$; this is shown by a fixed point argument essentially identical
to Lemma 2. The four quartic running coupling $g^i_h$ obeys to
equations of the form,if $i=(2o+),(2o-),(2p+),(2p-),(4+),(4-)$
$$g^i_{h-1}=g^i_{h}+\b^i_h+R_h^i\Eq(iioo)$$
where $\b^i_h$ is given by the sum of trees with no endpoints at scale $i$,
only $g_L^k$ propagators and no endpoints to which are associated
$\n_k,\d_k$; if  $\n_{h,+},\n_{h,-}, \d_{h,+},\d_{h,-}$ are 
$O(\bar g_h \g^{\th h})$ then, as in \S 5, $R_h^i\le C\bar g_h  g^{\th h}$.

The flow of the quartic running constants
is even simpler as the one in the $h=0$ case as 
$|\b^i_h|\le C\bar g_h  g^{\th h}$. 
This can be proved as in \S 6 introducing the following reference model,
with

$$\VV_L=\sum_{\o,\s} \int d\kk_1...\int d\kk_4 \d(\sum_i \e_i\kk_i)
[g_{2,\s}^o \psi^{+}_{\kk_1,\o,\s}\psi^-_{\kk_2,\o,\s}
\psi^+_{\kk_3,-\o,\s}\psi^-_{\kk_4,-\o,\s}+
g_{2,\s}^p \psi^+_{\kk_1,\o,\s}\psi^-_{\kk_2,\o,\s}
\psi^+_{\kk_3,-\o,-\s}\psi^-_{\kk_4,-\o,-\s}$$
$$+g_{4,\s}
\psi^+_{\kk_1,\o,\s}\psi^-_{\kk_2,\o,\s}
\psi^+_{\kk_3,\o,-\s}\psi^-_{\kk_4,\o,-\s}]\Eq(ll1)$$
In this reference model the interaction has five
independent parameters, instead of three as in the previous case.
We can analyze by RG the reference model and we get the couplings 
$\tilde g^p_{h,2,+}, \tilde g^p_{h,2,-},\tilde g^o_{h,2,+}, \tilde g^o_{h,2,-},
\tilde g^o_{h,4,+}, \tilde g^o_{h,4,-}$.
The proof that their values remains close to the initial value 
is essentially identical to the analysis in \S 6; \eq(b1)
is replaced by 
$$<\psi^+_{\kk_1,+,\s}\psi^-_{\kk_2,+,\s}\psi^+_{\kk_3,-,-\s}\psi^-_{\kk_4,-,-\s}>_T=$$
$$g_{-,-\s}(\kk_4) \{G^2_{-,-\s}(\kk_3)[g_{2,\s}^o <\psi^+_{\kk_1,+,\s}\psi^-_{\kk_2,+,\s} \r_{\kk_1-\kk_2,+,\s}>_T$$
$$+g_{2,-\s}^p <\psi^+_{\kk_1,+,\s}\psi^-_{\kk_2,+,\s} \r_{\kk_1-\kk_2,+,-\s}>_T$$
$$+g_{4,-\s} <\psi^+_{\kk_1,+,\s}\psi^-_{\kk_2,+,\s} \r_{\kk_1-\kk_2,-,-\s}>_T]$$
$$+\int d\pp [g_{2,\s}^o <\psi^+_{\kk_1,+,\s}\psi^-_{\kk_2,+,\s}
\psi^+_{\kk_3,-,\s}\psi^-_{\kk_4-\pp,-,\s} \r_{\pp,+,\s}>_T+\Eq(b1)$$
$$\int d\pp g_{2,-\s}^p <\psi^+_{\kk_1,+,\s}\psi^-_{\kk_2,+,\s}
\psi^+_{\kk_3,-,\s}\psi^-_{\kk_4-\pp,-,\s} \r_{\pp,+,-\s}>_T+$$
$$\int d\pp g_{4,-\s} <\psi^+_{\kk_1,+,\s}\psi^-_{\kk_2,+,\s}
\psi^+_{\kk_3,+,\s}\psi^-_{\kk_4-\pp,+,\s} \r_{\pp,+,-\s}>_T]\}$$
By using the Ward identities of \S 7 one gets that
$\tilde g^p_{h,2,+}, \tilde g^p_{h,2,-},\tilde g^o_{h,2,+}, \tilde g^o_{h,2,-},
\tilde g^o_{h,4,+}, \tilde g^o_{h,4,-}$
remain close to the initial value, and this implies that $\b_h^i$ is 
asymptotically vanishing. This means that in presence of a magnetic
field one has Luttinger liquid behaviour 
also with a attractive interaction; of course this will be true only if $\bar h$
is non vanishing, and it is $O(h^\a)$ for some constant $\a>0$
(in fact $\bar h$ is finite if $|p_F^{\s}-p_F^{-\s}|\not=0$). 
\vskip.5cm
\section(22, Appendix)
\* \sub(5.2all){\it Ultraviolet decomposition}

It is convenient to
introduce an ultraviolet cut-off $N$ by writing
$$g^{[1,N]}(\xx,\yy)=\sum_{n=0}^N g^{(n)}(\xx,\yy)\Eq(a1)$$
where
$$g^{(n)}(\xx,\yy)={1\over L\b}\sum_{\kk\in\DD} \hat f_{u.v.}(\kk)
h_n(k_0){e^{-i\kk(\xx-\yy)}\over -ik_0-\tilde t\cos k+\tilde t
\cos p_F}\Eq(a11w)$$
with $h_0(k_0)=\chi(k_0)$ and $h_n(k_0)=\chi(\g^{-n+1}k_0)-
\chi(\g^{-n}k_0)$; it holds that
$\lim_{N\to\io} g^{[1,N]}(\xx,\yy)=g^{(u.v.)}(\xx,\yy)$ and,
for any integer $K$
$$|g^{(n)}(\xx,\yy)|\le {C_K\over 1+(\g^n|\xx-\yy|)^K}\Eq(a2)$$
We define
$$V^{(0)}(\phi)=\lim_{N\to\io}\log {1\over\NN_0}\int P(d\psi^{[1,N]})e^{\VV(\psi^{[1,N]}+\phi) }\Eq(a3)$$
We can integrate iteratively scale by scale, and after the integration 
of the scales $N,
N-1,..,k+1$ we get
$$V^{(k)}(\phi)=
\lim_{N\to\io}\log {1\over\NN_k}\int P(d\psi^{[1,k]})e^{\VV(\psi^{[1,k]}+\phi) }\Eq(a4)$$
It is well known that  
$V^{(k)}$ can be written as sum over {\it trees} $\t$
similar to the ones in \S 3 
(see for instance [GLM] for the analysis of the ultraviolet problem
in the Hubbard model in any dimension)
each of them bounded by, if $m_v$ 
is the number of endpoints of type $U$ following the vertex $v$ on $\t$ 
$$C^n [\max(U,|\n|,|\d|)]^m \g^{-n(m-1)}\prod_v \g^{-(n_v-n_{v'})(m_v-1)}\Eq(a5)$$ 
One can have $m_v=1$ only if $v$ is a trivial vertex 
following the first non trivial vertices on $\t$; 
then the terms with $m_v=1$ 
correspond to {\it self-contractions} or {\it tadpoles};
note however that no divergence are associate to self-contractions as
$g^{(n,N)}(\xx,\xx)$ is bounded uniformly in $N$.
Consider then a generic tree with all the sets $P_v$ assigned; the 
simple expectations over the trivial vertices in the tree
with $m_v=1$ before the first non trivial vertex $\bar v$ can be 
explicitly computed, giving  
$\psi^{+(\le n_{\bar v})}_\xx\psi^{-(\le n_{\bar v})}_\xx g^{(n_{\bar v},M)}(\xx,\xx)$; 
the rest of the tree is bounded by an expression like \equ(a5)
with $m_v>1$, so that by 
summing over all the scales and the trees the bound \equ(2.62a)
is found.
\* \sub(5.2aokk){\it Spin symmetry}

Finally the symmetry property
\equ(2.62b) follows from the $SU(2)$ invariance of the Hubbard model.
A direct way to check this property consists in 
expanding the truncated expectations corresponding to the integration 
of $\psi^{u.v.}$
in terms of {\it Feynmann graphs}. The interaction 
can be also written in the following way,
making more explicit the spin symmetry of the Hubbard model 
$$\VV={U\over 2} \int dx_0 \sum_x (\sum_\s \psi^+_{\xx,\s}\psi^-_{\xx,\s})
(\sum_{\s'}  \psi^+_{\xx,\s'}\psi^-_{\xx,\s'})+\n \sum_{x,\s} \int dx_0 
\psi^+_{\xx,\s}\psi^-_{\xx,\s}+\d \sum_\s \int dx_0 \sum_{x,\s}
t_{x,y}\psi^+_{\xx,\s}\psi^-_{\yy,\s}
\Eq(m)$$
As usual, the Feymann graphs are obtained representing as vertices
the three addends in \equ(m) with four or two oriented half-lines, and contracting
in all possible ways the half lines with consistent orientation; it is also
convenient to represent the quartic term as a couple of two half-lines connected
by a wigghly line, representing the interaction.
The value of each Feynmann graph is obtained associating to each 
line a propagator $g^{u.v}(\xx;\yy)$ and integrating over all
the coordinates; the contributions from graphs with four uncontracted
half lines has
in general the form 
$$\int d\xx_1...\int d\xx_4 \psi^+_{\xx_1,\s}
\psi^-_{\xx_2,\s}\psi^+_{\xx_3,\s'}\psi^+_{\xx_4,\s'}W^0_{\s,\s'}(\xx_1,..,\xx_4)\Eq(m1)$$
In order to prove that the kernel is spin-independent,that is
$$W_{\s,\s}=W_{\s,-\s}\Eq(m2)$$
we note that in the Feynmann graph we can identify
a line of propagators $g^{u,v}(\xx,\yy)$
(possibly
a point) connecting 
$\psi^+_{\xx_1,\s}$ with $\psi^-_{\xx_2,\s}$, and another line
connecting 
$\psi^+_{\xx_3,\s'}$ with $\psi^-_{\xx_3,\s'}$; on such two lines 
there are points to which are attached  
wiggly lines 
to which are attached the fields
$\sum_{\s''}\psi^+_{\s''}\psi^-_{\s''}$; the crucial point is that such expression 
does not depend from the fact that it is connected by the wigghly line to a $\s$ or $\s'$ line.
Hence the contributions to
$W^0_{\s,\s}$ and $W^0_{\s,-\s}$
can possibly differ only
because in one case there is a line of propagators 
$\s$ and in the other case $-\s$; but the propagators are spin-independent
hence the values of such two contributions are identical (and independent from $\s$). 
The same argument can be repeated to prove that 
$W^h_{\s,\s}=W^h_{\s,-\s}$, by performing a single scale integration
with propagator $g^{\ge h}(\xx,\yy)$.

\baselineskip=12pt
\vskip1cm

\centerline{\titolo References}
\*
\halign{\hbox to 1.2truecm {[#]\hss} &%
\vtop{\advance\hsize by-1.25truecm\0#}\cr
A &{P.W. Anderson. The theory of superconductivity on high $T_c$ cuprates, Princeton University Press, Princeton 
(1997)}\cr
BM1& {G. Benfatto, V.Mastropietro.
{\it Rev. Math. Phys.} 13 (2001), no. 11, 1323--143}\cr
BM2& {G. Benfatto, V.Mastropietro.
{\it Comm. Math. Phys.}231, 97-134 (2002)}\cr 
BM3& {G. Benfatto, V.Mastropietro.
{\it Jour. Stat. Phys.} (2004)}\cr
BM4& {G. Benfatto, V.Mastropietro.
to appear in {\it Comm. Math. Phys.}}\cr 
FK& {H.Frahm,V.E.Korepin. {\it Phys Rev B} 42, 10553 (1990).}\cr
GLM& {G.Gallavotti, J.Lebowitz, V.Mastropietro. {\it Jour. Stat. Phys.}
108, n. 5-6, 831--861 (2002). }\cr
H &{F.D.M. Haldane.  {\it Phys. Rev. Lett.}45, 1358--1362 (1980).}\cr
L& {E.H.Lieb. The Hubbard model: its physiscs and mathematical physics, ed. D. Baeriswyl,
D. Campbell, J. Carmelo, F. Guinea, E.Luis, Nato ASI seroes, phys vol 343 (1995).}\cr 
LW& {E.H.Lieb, F.Y.Wu. {\it Phys. Rev. Lett.}20, 1445--1449 (1968).}\cr
ML&{D. Mattis, E. Lieb. {\it J. Math. Phys.} {\bf 6}, 304--312 (1965). }\cr
O& {A.A. Ovchinnikov. {\it Sov.Phys. JETP} 30,1160 (1970).}\cr 
OS& {M.Ogata, H.Shiba. {\it Phys. Rev. B} 41,2326 (1990).}\cr 
PS &{A.Parola S.Sorella. {\it  Phys.Rev. Lett.} 64, 1831-1834 (1990)        }\cr
%
%
RS& {A.Rosch, N.Andrei. {\it Phys Rev Lett} 85,5 1092--1096 (2000). }\cr
S & {J. Solyom. {\it Adv. Phys.} 28, 201--303 (1979).}\cr
T & {M. Takahashi. {\it Prog. Theor. Phys.} 89, (1972).}\cr
}

\bye

\bye

%% file: treelut.txt
{\ins{30pt}{85pt}{$r$}\ins{50pt}{85pt}{$v_0$}\ins{130pt}{100pt}{$v$}%
\ins{35pt}{-2pt}{$j$}\ins{55pt}{-2pt}{$j+1$}\ins{135pt}{-2pt}{$h_v$}%
\ins{215pt}{-2pt}{$-1$}\ins{235pt}{-2pt}{$0$}\ins{255pt}{-2pt}{$+1$}}%